%% file: main.tex
\documentclass{article}

\usepackage[final]{neurips2022}

\usepackage{tikz}
\usetikzlibrary{trees,shapes}

\usepackage[utf8]{inputenc} 
\usepackage[T1]{fontenc}    
\usepackage{hyperref}       
\usepackage{url}            
\usepackage{booktabs}       
\usepackage{amsfonts}       
\usepackage{nicefrac}       
\usepackage{microtype}      
\usepackage{xcolor,wrapfig}         
\usepackage{algorithm}
\usepackage{algpseudocode}
\usepackage{caption}
\usepackage{subcaption}
\usepackage{graphicx}

\title{What should clubs monitor to predict future value of football players}

\author{Ali Baouan\thanks{CMAP, CNRS, École polytechnique, Institut Polytechnique de
Paris, 91120 Palaiseau, France},\ \   Elsa Bismuth\footnotemark[1],\ \ Aurèle Bohbot\footnotemark[1],\ \ Sebastien Coustou\thanks{Parma Calcio 1913 Performance and Analytics},\\ 
Mathieu Lacome\footnotemark[2],\ \ Mathieu Rosenbaum\footnotemark[1]}

\begin{document}
\maketitle
\input{sections/abstract}
\input{sections/introduction}
\input{sections/methodology}

\input{sections/data}

\input{sections/results}
\input{sections/conclusion}
\vspace{-.8em}
\paragraph{Acknowledgement:}The authors thank Marc Mayor for inspiring discussions.
\clearpage
\newpage
\nocite{*}
\bibliographystyle{apalike}
\bibliography{references}

\input{sections/annex}

\end{document}

%% file: sections/abstract.tex
\begin{abstract}
Huge amounts of money are invested every year by football clubs on transfers. For both growth and survival, it is crucial for recruiting departments to make smart choices when targeting players. Therefore, it is very important to identify the right parameters to monitor to predict market value. The following paper aims at determining the relevant features that successfully forecast future value for football players. Success is measured against their market value from TransferMarkt. To select prominent features, we use Lasso regressions and Random Forest algorithms. Some obvious variables are selected but we also observe some subtle dependencies between features and future market value. Finally, we rank the Golden Boy nominees using our forecasts and show our methodology can successfully compare football players based on their quality.
\end{abstract}

%% file: sections/introduction.tex
\section{Introduction}
Various clubs such as Ajax Amsterdam, Borussia Dortmund, S.L. Benfica and FC Porto consistently maintain an excellent and competitive squad while frequently selling their most talented players. Combining squad turnover and quality requires a lot of work by the recruiting department to identify and correctly evaluate potential talent. The conventional scouting process typically follows manual procedures and requires substantial resources and time to spot suitable candidates for a position. Moreover, filtering the most relevant candidates manually from a large list of prospective ones is time-consuming and may lead to missing interesting targets. In such a competitive setting, the use of data science is becoming crucial and can give clubs new tools to improve their recruiting methodology.

To maintain financial sustainability, clubs are looking for players whose performance and market value are likely to increase over the following years. Therefore, they try to determine parameters correlated with the improvement of the players' quality. The problem of the identification of the relevant predictors of market value is well described in the literature. Numerous factors such as age, nationality, and effort are explored in \cite{majewski2016identification,wicker2013no} and can be classified into three categories - player characteristics, player
performance, and player popularity. In this paper, we focus on player performance along with specific characteristics such as age, height, quality of youth formation, and the quality of the league. Moreover, we are interested in finding signals correlated with future value rather than the present one. The goal is to evaluate, for each position on the pitch, the effects of these factors on a player's future market value. 

The performance data we use is retrieved from Wyscout. It consists of 111 in-game statistics for 36,882 players. Player characteristics and market value are collected from \href{https://www.transfermarkt.com/}{www.transfermarkt.com}.  \href{https://www.transfermarkt.com/}{TransferMarkt}, the leading website on the football transfer market, provides an independent valuation of players. The market values on the website are computed using different valuation methods. An important factor is the TransferMarkt community, whose members discuss and subjectively evaluate the market values of players. \cite{herm2014crowd} provides a detailed explanation of how the Transfermarkt crowd evaluation works. Since the football transfer market is highly illiquid, the TransferMarkt values are among the best publicly available estimates. In particular, \cite{he2015football} shows the economical valuation of TransferMarkt closely matches the prices
paid for the transferred players.

In this work, we consider a player's market value as an approximative representation of in-game quality, neglecting popularity aspects in the valuation of players. Using performance statistics of football players over two years, we implement regressions to predict their market value two years after their last performance date. The aim is to determine, separately for each position on the pitch, the relevant variables that help predict future market value.  Given the large number of available features and the fact that they can be highly correlated, it is crucial to eliminate the redundant ones. In this regard, we use Lasso regressions and the Random Forest approach. Lasso regressions are well known for sparsifying the feature space while Random Forests provide a mechanism to evaluate feature importance and allow for non-linear interplay between predictors. As a second step, we analyze the resulting prominent features given by both approaches for each position and discuss their interpretability. Finally, we apply our forecasting methodology to rank the sixty nominees for the Golden Boy 2022 award and compare our result with the ranking of the jury.

In section \ref{sec:methodology}, we introduce the Lasso and Random Forest approaches and the methodology for selecting features. Section \ref{sec:data} describes the data we use and how it is processed. In section \ref{sec:results}, we present and discuss the results of our approach. The selected performance metrics show promising predictive power. For example, we are able to explain up to 60$\%$ of the variability of players' market value. Moreover, we observe that the Random Forest algorithm and Lasso regression perform similarly. Both approaches successfully predict the range of the future value and more importantly provide a list of relevant features specific to each position. Unsurprisingly, the playing time, age, and league average value stand out as important features for all positions, with the league's average value contributing to 20\% of the explained variance. Most of the remaining selected features are consistent with the needs of each position. We also observe surprising results, such as the negative impact of playing long balls on the future market value. Subsection \ref{sec:feat} presents a list of such predictors along with their interpretation. In section \ref{sec:youngplayers}, we establish empirical proof that our methodology holds more value in the evaluation of player quality compared to using the present market value. Our predicted top thirteen young players successfully includes nine out of the top ten published by the Golden Boy award. Moreover, the ranking based on our prediction achieves a significant positive correlation with the jury ranking.

%% file: sections/methodology.tex
\section{Methodology}
\label{sec:methodology}

For finding the key features that help predict the player value at a given horizon, we use two different approaches: Lasso regression and Random Forest. In particular, we compare their accuracy in the prediction and the selected features. In this section, we describe briefly these two methodologies.
\subsection{Lasso regression}
The Lasso (Least Absolute Shrinkage and Selection Operator) method, or $L_1$  regularization, is introduced in \cite{tibshirani1996regression}. Suppose we have data $(x^i, y_i)_{1\leq i\leq n}$ where $x^i=(x_{i1},x_{i2},\dots,x_{id})$ are the predictor variables and $y_i$s are the responses to predict. We further assume that the $(x_{ij})$ are standardized.
 
 The Lasso estimate $(\hat{\alpha},\hat{\beta})$ is defined by: 
\[     \hat{\alpha},\hat{\beta} =  \textup{arg}\min \left [ \sum_{i=1}^{n}  \left ( y_i - \alpha - \sum_{j=1}^{d} \beta_j x_{ij} \right ) ^2\right ]  \; \textup{ subject to } \; \sum_{j=1}^{d} |\beta_j| \leq t .
\]

Equivalently, the Lasso estimate can be written under the langrangian form without a constraint: 
\[     \hat{\alpha},\hat{\beta} =  \textup{arg}\min \left [ \sum_{i=1}^{n}  \left ( y_i - \alpha - \sum_{j=1}^{d} \beta_j x_{ij} \right ) ^2 +\lambda \sum_{j=1}^{d} |\beta_j| \right ] . 
\]
 
    This method is particularly relevant in our setting. Compared to the standard linear regression, Lasso can be used when we have a large number of highly correlated predictors, even when we have more features than datapoints $d>n$. Furthermore, Lasso solutions are particularly sparse. For a large enough value of $\lambda$, features with a limited influence on the target are assigned a null coefficient.
\subsection{Random Forest algorithm}
    Random Forest algorithms are introduced in \cite{breiman2001random}. They are based on a standard approach called \textit{decision tree}. We provide a brief explanation of how this algorithm works in the following section.
\subsubsection{Decision trees}
    The decision tree algorithm is an approach that uses a set of binary rules to calculate a target class or value. They aim at dividing the feature space into multiple partitions recursively and then assigning to each cell of the partition the average target of its samples. For each new node, a feature/split is optimally chosen to maximize a performance metric (entropy for classification and variance for regression). Each decision tree has three levels:
    \begin{enumerate}
        \item Root nodes: entry points of a dataset.
        \item Inner nodes: a collection of binary tests that divide the feature space or subspace.
        \item Leaf nodes: target prediction, usually the average of the targets of samples from the training set that reach the leaf node. Sometimes, a linear regression is fitted on the samples of each leaf node.
    \end{enumerate}
\tikzset{
  breakarrow/.style={->, dashed},
  myarrow/.style={solid, ->},
  varnode/.style = {solid,shape=rectangle, rounded corners,
    draw, align=center,
    top color=red!20, bottom color=white},
resnode/.style = {solid,shape=rectangle, rounded corners,
    draw, align=center,
    top color=green!20, bottom color=white,scale=0.99},
    violet/.style = {shape=rectangle, rounded corners, draw, align=center, top color=white, bottom color=blue!20}}

\begin{figure}[h!]
\centering
\begin{tikzpicture}[level distance=1.5cm,
level 2/.style={sibling distance=7.2cm},edge from parent fork down,
level 3/.style={sibling distance=3.7cm},level 4/.style={sibling distance=1.8cm}]
  \node [varnode] {Datapoint \\ $x_1$: Logarithm of the league average value\\$x_2$: Total minutes played \\ $y$: Logarithm of the future market value}
  child { node[violet] {$x_1$ $<15.75$}   
  child { node[violet] {$x_1$ $<13.25$}  
    child { node[violet] {$x_2$ $<970$}  
        child  { node  [resnode] {$\hat{y}$=12.60} edge from parent node[above,draw=none] {True} }
        child { node  [resnode] {$\hat{y}$=13.24} edge from parent node[above,draw=none] {False} }
            edge from parent node[above,draw=none]{True}}
    child { node[violet] {$x_2$ $<790$} 
        child  { node [resnode]  {$\hat{y}$=13.21} edge from parent node[above,draw=none] {True} }
        child  { node [resnode]  {$\hat{y}$=14.13} edge from parent node[above,draw=none] {False} }
            edge from parent node [above,draw=none]{False} }
  edge from parent node[above,draw=none]{True}}
  child { node [violet]{$x_2$ $<2120$}   
    child { node [violet]{$x_2$ $<450$}  
        child  { node [resnode]  {$\hat{y}$=12.22} edge from parent node[above,draw=none] {True}}
        child  { node [resnode]  {$\hat{y}$=15.02} edge from parent  node[above,draw=none] {False}} 
          edge from parent node[above,draw=none]{True}}
    child { node[violet] {$x_2$ $<3100$} 
        child  { node  [resnode] {$\hat{y}$=16.15} edge from parent node[above,draw=none] {True}}
        child  { node [resnode]  {$\hat{y}$=17.24} edge from parent node[above,draw=none] {False}}
          edge from parent node[above,draw=none]{False}}
    edge from parent node[above,draw=none]{False}}
    edge from parent node[above,draw=none]{}};
\end{tikzpicture}
\caption{An example of binary tree to predict a player's value using his league's average value and his total minutes played. The feature space is partitioned using the conditions at each node. A prediction is decided at each leaf, if all criteria along its path are satisfied.}
\label{fig:sortingtree}
\end{figure}
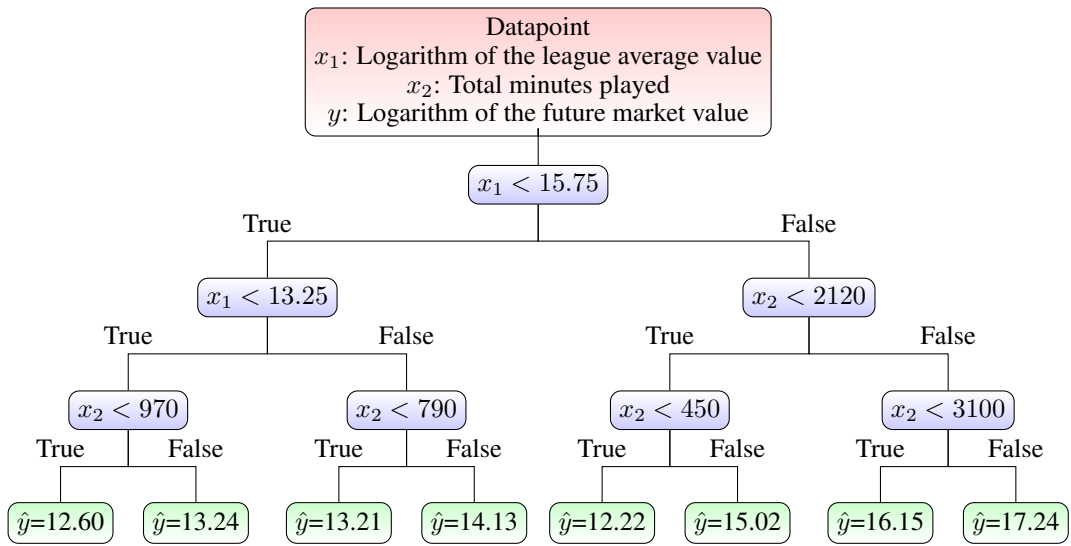

\begin{figure}[t!]
    \centering
    \begin{minipage}{0.95\textwidth}
        \centering
        \scalebox{0.8}{\includegraphics{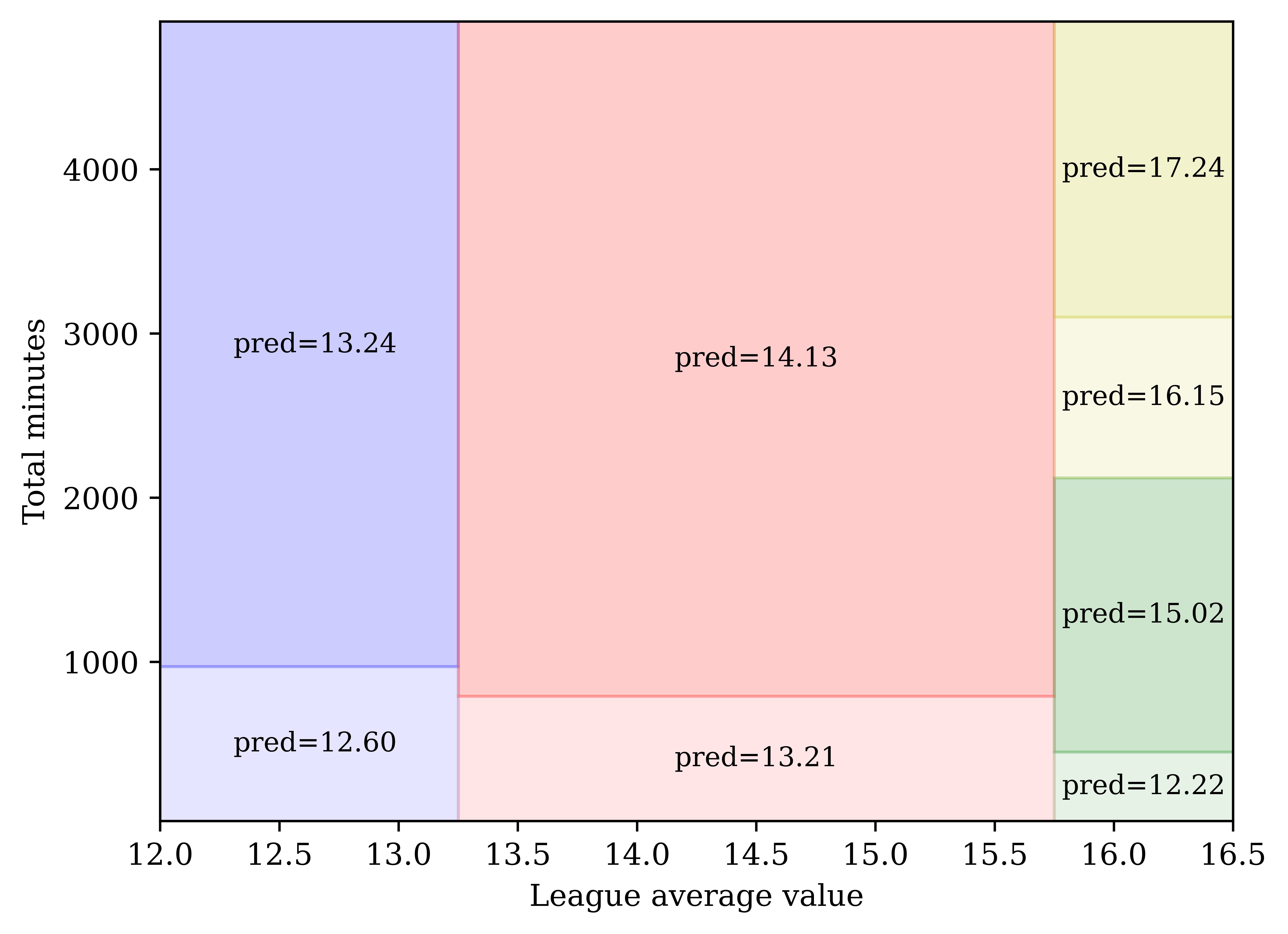}}\vspace*{-.5em}
        \captionof{figure}{Partition of the feature space given by the decision tree in Figure \ref{fig:sortingtree} .}\label{fig:partitionspace}      
    \end{minipage}
\end{figure}

For example, if we try to predict the log-value of a player given as inputs the average value in his league and his playing time, a binary tree can be constructed based on information from each feature. Figure \ref{fig:sortingtree} shows an example of such decision tree. At each node level, a logical test is performed on the input features to determine the path of the data point in the graph. The attained leaf node determines the predicted value. Generally, this prediction is the average of targets of samples reaching this leaf during training. The collection of binary tests can be viewed as a way to cluster data in rectangular blocs where each new data point is assigned as a prediction the mean of targets of training points in his cluster, see Figure \ref{fig:partitionspace} for an example.

Algorithm \ref{alg:decisiontree} describes how a decision tree is fitted to a dataset. The algorithm is computationally fast and depends on a stopping criterion, generally a minimal number of samples reaching each leaf or a maximum depth. Though they offer an intuitive way to interpret data, decision trees are prone to overfitting. In fact, decision trees usually have low bias and high variance. This can be seen if we consider extreme cases of the stopping criterion. A complex enough decision tree can divide the input space into subspaces containing each individual training data point, and therefore memorizing the dataset rather than learning a dynamic relationship between features and the target.
    
    \begin{algorithm}[h!]
\caption{Regression trees - CART algorithm for building a decision tree.}\label{alg:decisiontree}
\begin{algorithmic}[1]
\State {} Consider the root node N with all input data $D$.  
\State {} For each feature F and value T that splits the samples assigned to N into subsets $D_{true}$ and $D_{false}$, compute the impurity, i.e the weighted average of variance of targets in each subset.
\State {} Choose the couple (F,T) that minimises the resulting impurity. 
\State Attach nodes $N_{true}$ and $N_{false}$ to parent node N. 
\State Repeat steps 2-4 until a stopping criterion is met.
\end{algorithmic}
\end{algorithm}

    \subsubsection{Random Forests}
    Random Forests belong to the family of ensembling techniques. It consists of a collection of decision trees where randomness is injected into several parts of the building process. As can be seen in equation \ref{eq:RFpred}, the Random Forest output is the mean of the predictions of the individual trees. Decision trees are ideal candidates for such methods since they exhibit low bias and high variance and can benefit from averaging. The use of Random Forest can lead to significant improvements in out-of-sample accuracy in comparison to a single binary tree.
    \begin{equation}
    \label{eq:RFpred}
        \hat{y}_{RF}=\frac{1}{n (\textup{\small trees})} \sum_{k=1}^{n (\textup{\small trees})} \hat{y}_{\textup{\small tree}_k} .
    \end{equation}
    Out of the numerous randomization procedures, bagging has proved its efficiency in multiple applications. Bagging consists of combining estimators trained on random bootstrap samples of the dataset. \cite{breiman2001random} combines bagging with random variable selection at each node, where only a subset of features is considered for each new split. This makes sense when the number of features is large and the target can be explained in different ways using the inputs. This might lead to a slight increase in bias, which is generally compensated by the decrease of variance due to aggregation.

    \subsubsection{Feature importance}
    In most machine learning tasks and particularly for our work, it is important to determine the input features that are most important to predict the target. Random Forests provide a mechanism to evaluate the importance of a feature, therefore making the methodology more interpretable.
    
    A feature's importance in a decision tree is computed as the average importance of the nodes where the feature is involved. A node $\textup{n}$'s importance is defined as the decrease of impurity in its child nodes: 
    \[
        \textup{importance}(\textup{n})=  w_{\textup{n}} V_{\textup{n}} - \sum _{\textup{n' child of n}} w_{n'} V_{n'}.
   \]
    
    where \begin{itemize}
        \item $w_{\textup{n}}$: is the fraction of samples that reach node $\textup{n}$
        \item $V_{\textup{n}}$: is the variance of the target of samples that reach node $n$
    \end{itemize}
    The importance of the feature is then calculated :  
    \[
        \textup{importance}(\textup{feature})= \frac{\sum\limits_{\textup{n : node that splits in feature}}\textup{importance}(\textup{n})}{\sum\limits_{\textup{n : node }} \textup{importance}(\textup{n})}.
    \]

A feature's importance in a Random Forest can be defined as the sum of its importance in all the trees it's sampled in.

%% file: sections/data.tex
\section{Data}
\label{sec:data}
\subsection{Data description}
The data we use is made of a dataset retrieved from Wyscout, an Italian company that collects data on football matches, and two datasets from TransferMarkt.
    
    In the Wyscout dataset, each row consists of information on a player's performance in a given game. It includes an ID number that can be matched with the TransferMarkt ID, the date of the match as well as 111 in-game statistics, all of which can affect the market value of the player, see Tables \ref{tab:wyscoutglos1}, \ref{tab:wyscoutglos2} and  \ref{tab:wyscoutglos3} for a list of the provided statistics. In total, there are 2,646,549 games of 36,882 players across 45 leagues between 2015 and March 2022. The list of considered leagues is as follows: 
    \begin{itemize}
        \item First division of: Italy, England, Spain, Germany, Turkey, France, Holland, Poland, Russia, Belgium, Portugal, Greece, Scotland, Austria, Switzerland, Croatia, MLS, Cyprus, Brazil, Romania, Denmark, Sweden, Serbia, Norway, Ukraine, Czech Republic, Japan, Argentina, China, Uruguay, Korea, Colombia, Georgia, Chile, Iceland, Paraguay, Montenegro.
        \item Second division of: England, Italy, Spain, Germany, France, Portugal, Brazil.
        \item Third division of: England.
    \end{itemize}

    The first TransferMarkt dataset gives information on players' value at different recorded dates. It contains the player's ID, the player's name, his date of birth, the date of the value, and the recorded TransferMarkt value. In total, there are 415,890 values for 33,439 players between 2000 and 2022.

    In the second TransferMarkt dataset, each row gives general information about a player such as his name, position, club, league, social medias and other pieces of information about his identity. It also includes their ID, which allows us to combine the 3 datasets. In this dataset, there are 21,898 players and 26 features. 

\subsection{Preprocessing}
    
    The three datasets have 12,133 players in common. The goal is to predict player values in a two year horizon. Thus, each player represents a data point in the regressions and we try to forecast his last available Transfermarkt value using performance data from Wyscout over the window between 730 days and 1460 days before the value date. Match statistics are summed over this window and divided by the total game time for each player to work with metrics per unit of time. Several additional features are added: 
    \begin{itemize}
        \item To account for non-linearities: ratio of successful statistics over attempted ones. Table \ref{tab:feateng} in the annex provides a list the engineered features.
        \item  Also to account for non-linearities: we added the squares of age, height, average goals per minute, average assists per minute, average shots per minute
        \item The average league value: computed by averaging the TransferMarkt market value of all players belonging to the league.
        \item A boolean feature taking as value 1 if the player's last youth team is in the 20 best youth systems and 0 otherwise.
    \end{itemize} 

The features are normalized to feed a \textit{flat} dataset to the Lasso regression. To keep homogeneity, we take the logarithm of the average league value and we substract its mean. For the rest of the continuous features, the normalization scheme is as follows:
    
\[
x_{i,j} = \frac{x_{i,j} - \overline{x_j}}{\max(x_j)}.
\]
where $\overline{x_j} = \sum\limits_{i=1}^n \frac{x_{i,j}}{n}$ is the average of the $j^{th}$ feature and $\max(x_j) = \max_{i} x_{i,j}$ its maximum value.

    A player's position provides crucial information as each different position has a different average value, see Table \ref{tab:posval}. We can either add the position as a feature in the dataset or perform separate regressions for the data points of each position. The latter seems more relevant since we can isolate precise features specific to each different position. Furthermore, the number of data points per position is reasonably high, see Table \ref{tab:numplaypos}. The dataset is divided into 8 parts depending on the position according to the TransferMarkt terminology (a player can appear in multiple positions):
    \begin{itemize}
        \item GK: Goalkeepers
        \item FB: Left Back and Right Back
        \item CD: Central Back and Sweeper
        \item CDM: Defensive Midfielder
        \item MD: Central, Right and Left Midfielder
        \item AM: Attacking Midfielder
        \item WG: Left Winger and Right Winger
        \item FWD: Centre-Forward
    \end{itemize}
\begin{table}
\subfloat[Average market value per position.]{
 \begin{tabular}{lr}
        \toprule
        \textbf{Position} &  \textbf{Average value} \\
        \midrule
        Goalkeeper   &   1,672,808 \$ \\
        Fullback &   2,799,533 \$ \\
        Central defender    &   3,044,609 \$ \\
        Defensive midfielder  &   3,620,264 \$ \\
        Central midfielder     &   3,466,659 \$ \\
        Attacking midfielder  &   4,059,712 \$ \\
        Winger    &   3,764,969 \$ \\
        Attacker     &   3,910,247 \$ \\
        \bottomrule
        \end{tabular}
        \label{tab:posval}
}%
\hfill
\subfloat[Number of datapoints per position.]{
\begin{tabular}{  c  c} 
\toprule
  \textbf{Position} &  \textbf{Number of players}  \\ 
  \midrule
  Goal    &               1227 \\
        Fulldef &               3055 \\
        Cdef    &               2759 \\
        Defmid  &               2365 \\
        Mid     &               4112 \\
        Attmid  &               2193 \\
        Wing    &               3147 \\
        Att     &               2448 \\
  \bottomrule
\end{tabular}\label{tab:numplaypos}\vspace{0.2em}
}    
 \caption{Market value and number of datapoints per position.}
\end{table}

\begin{figure}[t!]
    \centering
    \begin{minipage}{0.47\textwidth}
        \centering
        \scalebox{0.46}{\includegraphics{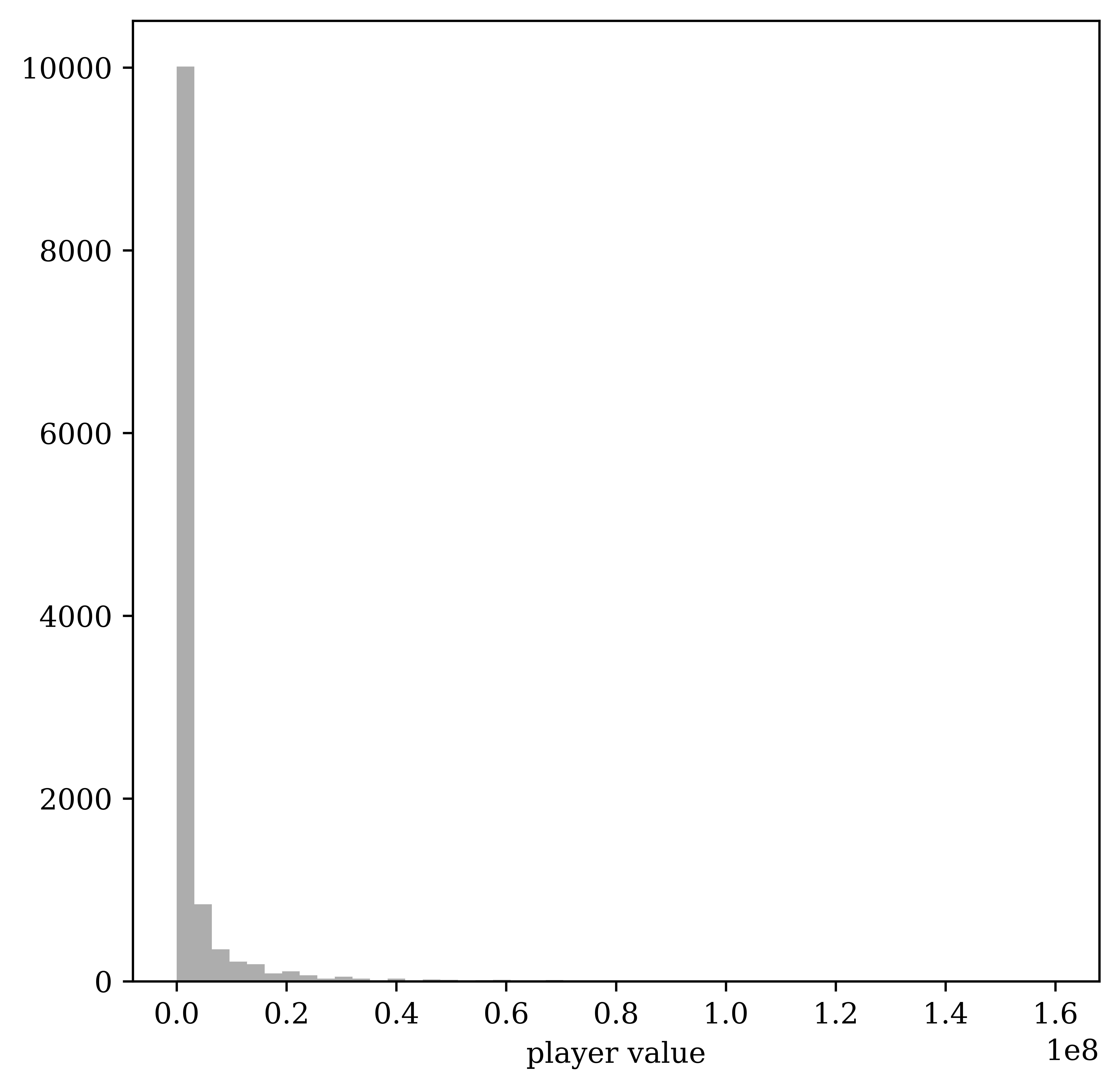}}\vspace*{-.5em}
        \captionof{figure}{Player market value distribution.}\label{fig:dist1}      
    \end{minipage}
    \begin{minipage}{0.47\textwidth}
        \centering
        \scalebox{0.46}{\includegraphics{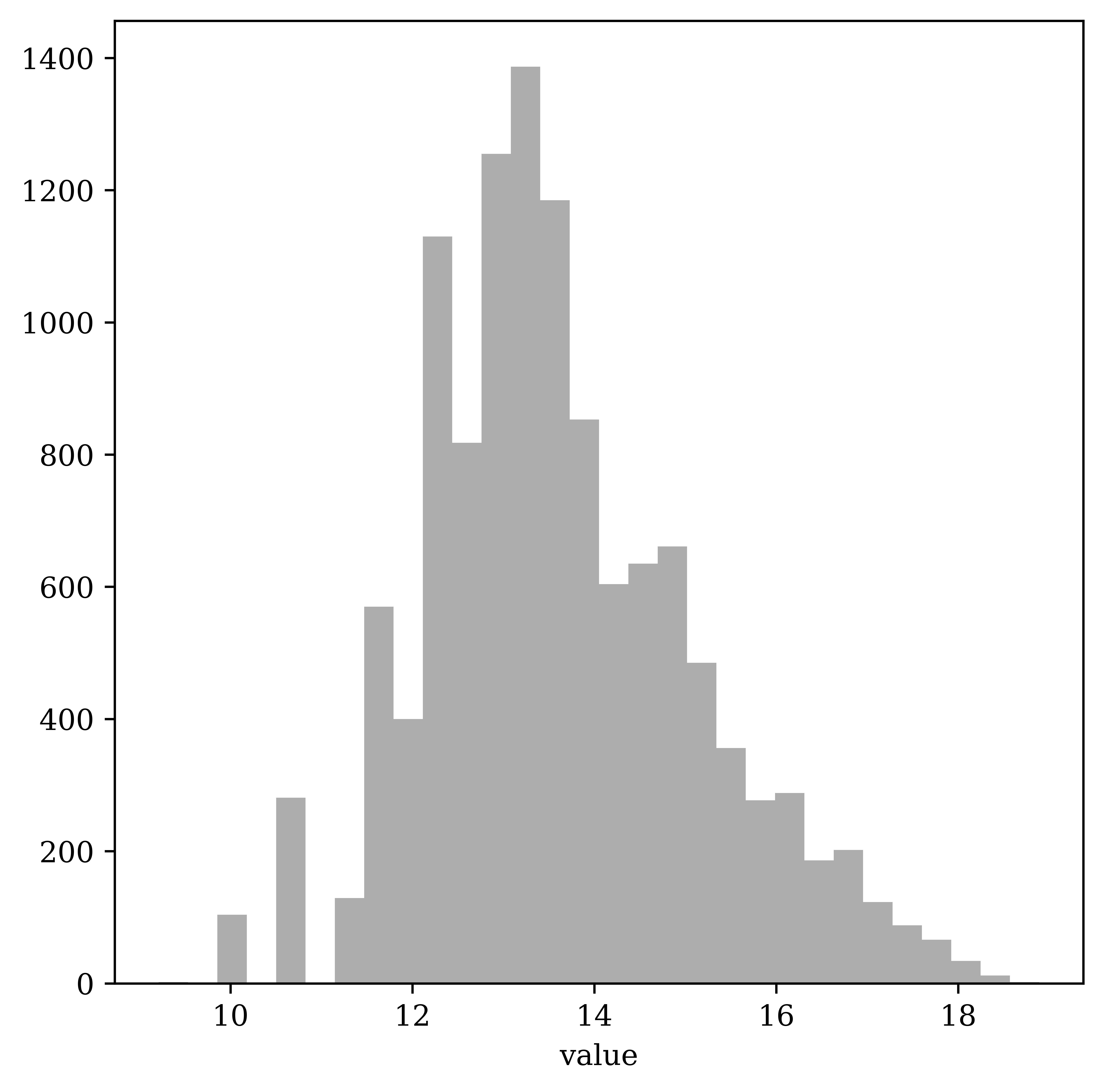}}\vspace*{-.5em}
        \captionof{figure}{Player market log-value distribution.}\label{fig:dist2}      
    \end{minipage}
\label{fig:valuesdistlog}
\end{figure}

    Concerning the market value in two years, its distribution is widely spread and skewed. A way to overcome this issue is to consider the logarithms of the player values. After this operation, the target values lie between 0 and 20, see Figures \ref{fig:dist1} and \ref{fig:dist2}.

%% file: sections/results.tex
\section{Results and discussion}
\label{sec:results}
In this section, we discuss the results of the prediction of players' value in a two year horizon.
\subsection{Validation metrics}
For the validation process, we use k-fold cross validation ($k=5$). We use mean squared error (MSE) and cross-validated R-squared ($\textup{R}^2)$ to compare our approaches and choose hyper-parameters. Cross-validated R-squared is computed using the average of the out-of-sample errors in each fold of the cross-validation: 

\[
    \textbf{R}^2 = 1-\frac{MSE_{\textup{cross-validation}}}{VAR(y)}.
\]
\paragraph*{Lasso Regression}

First, we fit the Lasso regression on the dataset of each position. The cross-validation score enables us to both evaluate the regression and choose an optimal Lasso parameter $\lambda$.

However, because the number of features is too important when using the optimal $\lambda$, usually exceeding thirty, we arbitrarily set $\lambda$ (between 0.004 and 0.01) to retrieve 10 to 15 features. This does not affect the cross-validation score much as long as we do not set very large values of $\lambda$.

\paragraph*{Random Forest}
A grid search on parameters encourages us to keep simple Random Forest parameters. We set the number of trees at 100 and the maximum depth of the tree at 6, as a good balance between accuracy and parsimony.
\subsection{Predicting player value two years later}

\subsubsection{Results}

We perform two different regressions for each approach and position dataset. A first regression includes the average league value as a feature, see Table \ref{tab:resulttable1}. This is a powerful predictor that explains a lot of the variance since it gives the estimators a first indication of the range of the player value. A second regression is performed on the dataset without this explanatory variable, see Table \ref{tab:resulttable2}.

We observe that suitably chosen performance metrics actually have predictive value. They can explain about 35\%  to 40\% of the future market value variance. The average league value adds a lot of information which results in an extra 20\% in $\textup{R}^2$.

Furthermore, we can see that the results obtained from the Lasso regression and the Random Forest are comparable in this dataset. This may be surprising since Random Forests have a lot more freedom in manipulating the input features. This can be explained by the limited amount of data that leads us to choose simple Random Forests. Moreover, the lasso estimator can also take advantage of the engineered non-linear features.

In this work, the part of the market value coming from other factors than in-pitch performance is considered noise. Therefore, our prediction should be seen as a \textit{filtered} market value where all other price factors are averaged out.

\begin{table}[t!]
        \centering
        \begin{tabular}{l|l l l|l l}
     
 {} & \multicolumn{3}{|c|}{\textbf{Lasso}} & \multicolumn{2}{|c|}{\textbf{RF}} \\
 \hline
\toprule
\textbf{Position} &     $\lambda$ &       \textbf{mse} &       \textbf{R}$^2$  & \textbf{mse} & \textbf{R}$^2$\\
\midrule
Goal      & 0.01 &1,074 & 48,5\% & 0,962 & \textbf{53,8}\% \\ 
Fullback   & 0.0055    &0,854 & \textbf{55,4}\% & 0,877 & 54,1\% \\ 
Central defender &   0.005     &0,865 & 57,7\% & 0,857 & \textbf{58,1}\% \\ 
Defensive Midfielder  &0.005      &0,857 & 60,0\% & 0,831 & \textbf{61,2}\% \\ 
Central Midfielder   &0.007    &0,814 & \textbf{61,5}\% & 0,814 & \textbf{61,5}\% \\ 
Attacking Midfielder  & 0.006      &0,982 & \textbf{55,9}\% & 1,050 & 52,8\% \\ 
Winger    &0.005    &0,915 & \textbf{57,4}\% & 0,980 & 54,3\% \\ 
Attacker    &0.005    &0,915 & \textbf{58,0}\% & 0,928 & 57,4\% \\ 

\bottomrule
\end{tabular}
\vspace{0.5em}
        \caption{Cross-validation accuracy results for Lasso and Random Forest (RF) when using average league value as a feature.}
        \label{tab:resulttable1}
    \end{table}

 \begin{table}[t!]
        \centering
        \begin{tabular}{l|l l l|l l}
     
 {} & \multicolumn{3}{|c|}{\textbf{Lasso}} & \multicolumn{2}{|c|}{\textbf{RF}} \\
 \hline
\toprule
\textbf{Position} &     $\lambda$ &       \textbf{mse} &       \textbf{R}$^2$  & \textbf{mse} & \textbf{R}$^2$\\
\midrule
Goal      & 0.01 &1,314 & 36,77\% & 1,173 & \textbf{43,7}\% \\ 
Fullback   & 0.0055    &1,154 & 39,6\% & 1,154 & \textbf{39,7}\% \\ 
Central defender &   0.005     &1,239 & \textbf{39,39}\% & 1,266 & 38,1\% \\ 
Defensive Midfielder  &0.005      &1,329 & \textbf{37,9}\% & 1,378 & 35,67\% \\ 
Central Midfielder   &0.007    &1,326 & 37,2\% & 1,318 & \textbf{37,6}\% \\ 
Attacking Midfielder  & 0.006      &1,378 & \textbf{38,7}\% & 1,434 & 35,5\% \\ 
Winger    &0.005    &1,352 & \textbf{37,1}\% & 1,389 & 35,3\% \\ 
Attacker    &0.005    &1,352 & 37,9\% & 1,291 & \textbf{40,6}\% \\ 

\bottomrule
\end{tabular}
\vspace{0.5em}
        \caption{Cross-validation accuracy results for Lasso and Random Forest (RF) without using average league value as a feature.}
        \label{tab:resulttable2}
    \end{table}

\subsection{Important features}
\label{sec:feat}
\subsubsection{General comments}

We can retrieve a pattern and see that some features have the same impact for all positions. We are going to discuss these features in this section.

\begin{figure}[t!]
    \centering
    \begin{minipage}{0.95\textwidth}
        \centering
        \scalebox{0.46}{\includegraphics{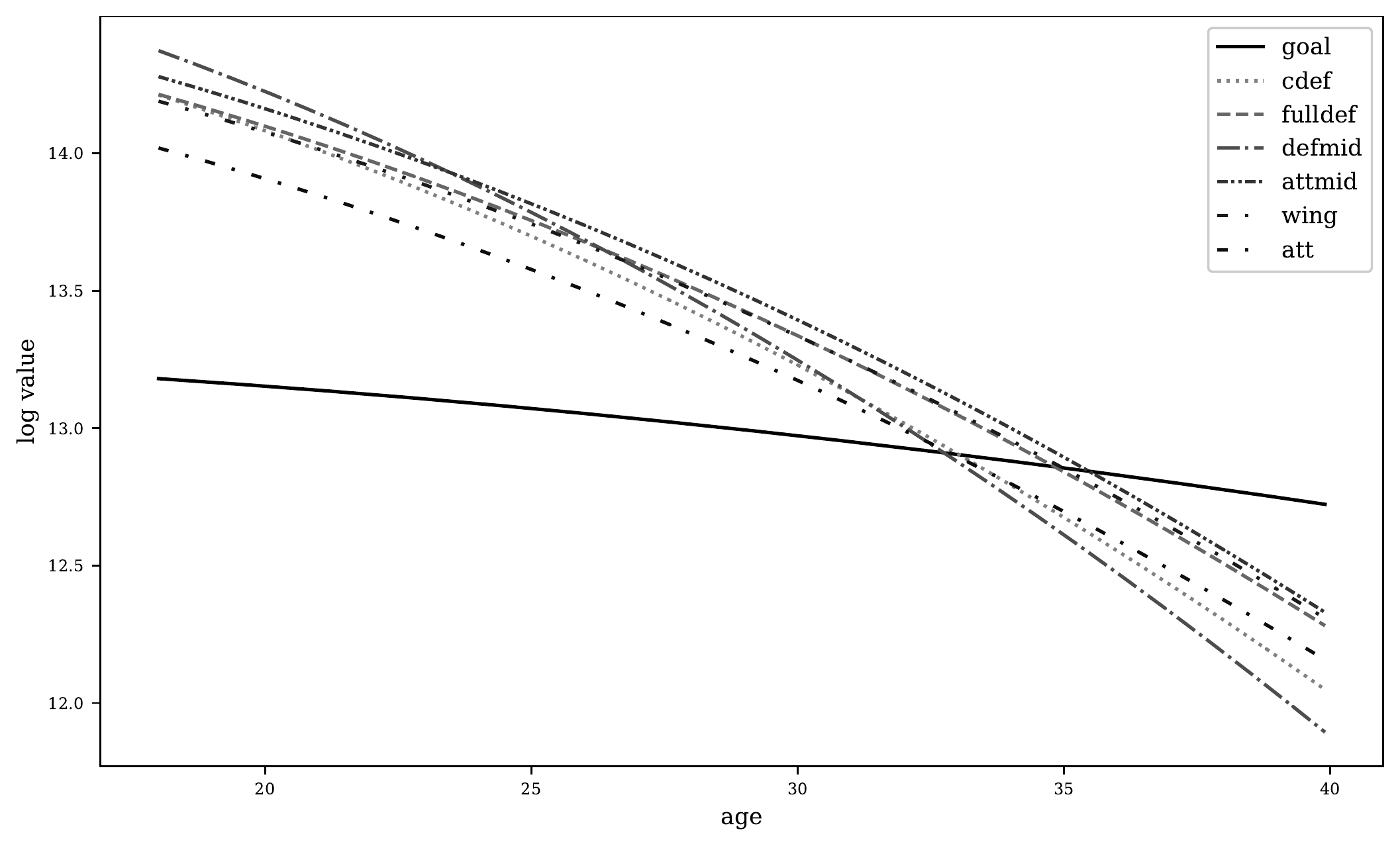}}\vspace*{-.5em}
        \captionof{figure}{Dependence of log value with respect to age.}\label{fig:dependenceage1}      
    \end{minipage}
    
\end{figure}

\begin{figure}[t!]
    \centering
\begin{minipage}{0.95\textwidth}
        \centering
        \scalebox{0.46}{\includegraphics{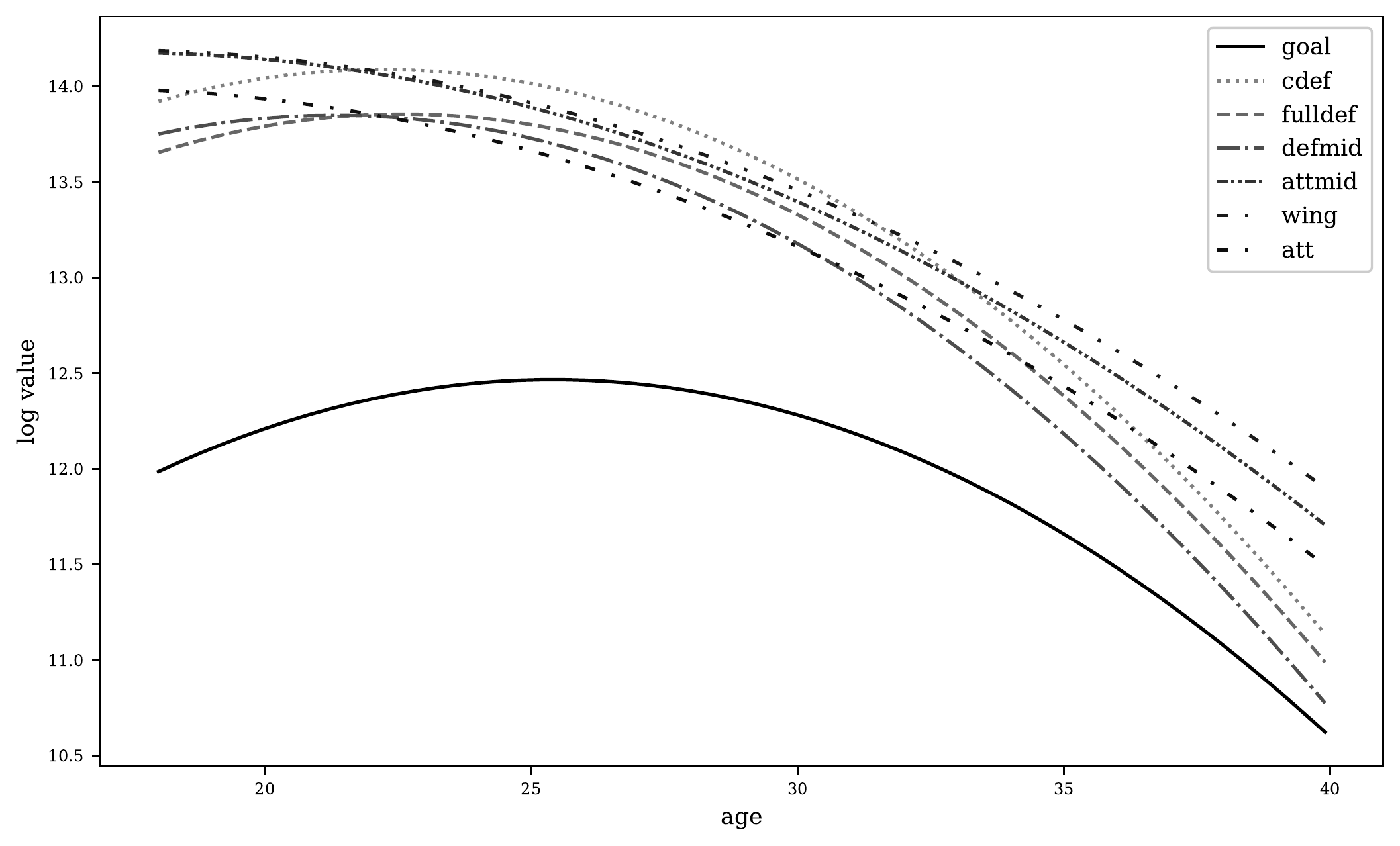}}\vspace*{-.5em}
        \captionof{figure}{Dependence of log value with respect to age when Lasso parameter is set to $\lambda=0,0001$.}\label{fig:dependenceage2}      
    \end{minipage}
\end{figure}

Figure \ref{fig:dependenceage1} represents the dependence of the log-value with respect to the age captured by the Lasso regression when all the standardized features are set to zero except the age and its square. The player's age has an overall negative effect on the value. Unsurprisingly, the rate of decrease is significantly lower for goalkeepers as they typically have longer careers. Since we set large values of $\lambda$, the Lasso regression only keeps the \textit{age\_squared} feature and assigns a negative coefficient to it. Therefore, in the range of values of the age variable, we only observe the decreasing part of a parabola. Figure \ref{fig:dependenceage2} shows the dependence of the log-value with respect to the age when using a small value for $\lambda$. In this case, the coefficient of the \textit{age} feature is positive and the regression captures the complex relationship between the market value and age with a bell-shaped parabola. In particular, a player's value should increase until his prime, as he accumulates experience and reputation. Furthermore, Lasso places the prime age for goalkeepers later than the other positions, confirming our previous interpretation.

\paragraph*{Random Forest}

We can see from the feature importance analysis of the Random Forest algorithm that the 4 most prominent features are almost always the same for every position, in this order of importance: $league\_avg\_value$, $total\_minutes\_on\_field$, and $age$.
It seems logical that these features are crucial. The league average value is always the top feature as it gives the algorithm a benchmark of what the value is in the league and is therefore useful. Then comes the $total\_minutes\_on\_field$, which proves that the time spent on the field by the player has a high influence on his value.
\paragraph*{Lasso Regression}

Looking at the features selected by the Lasso estimator, we recover the same important variables as Random Forests. This confirms the significant role they play in determining the value of the player. As expected, 
$total\_minutes\_on\_field$ and $league\_avg\_value$ have the highest positive coefficients, and $age\_sq$ the most negative coefficient.

Another feature that always stands out is $is\_top\_20$, which specifies whether the player's last youth club is one of the top 20 formation centers. It does not have a high coefficient but helps the regressions correct their valuation for certain players.

In the following subsections, are presented the most important features for each position with their interpretation. The displayed features are selected from the regression without average league value as a predictor.

\subsubsection{Goalkeepers}

Unsurprisingly, the features created specifically for goalkeepers only have an influence on goalkeepers. Furthermore, Random Forest gives more importance to minutes played compared to other positions. This is because goalkeepers are almost always either starters or full-time substitutes, which makes game time an even more crucial predictor to determine the value for goalkeepers.

\begin{figure}[t!]
    \centering
    \begin{minipage}{0.43\textwidth}
        \centering
        \scalebox{0.4}{\includegraphics{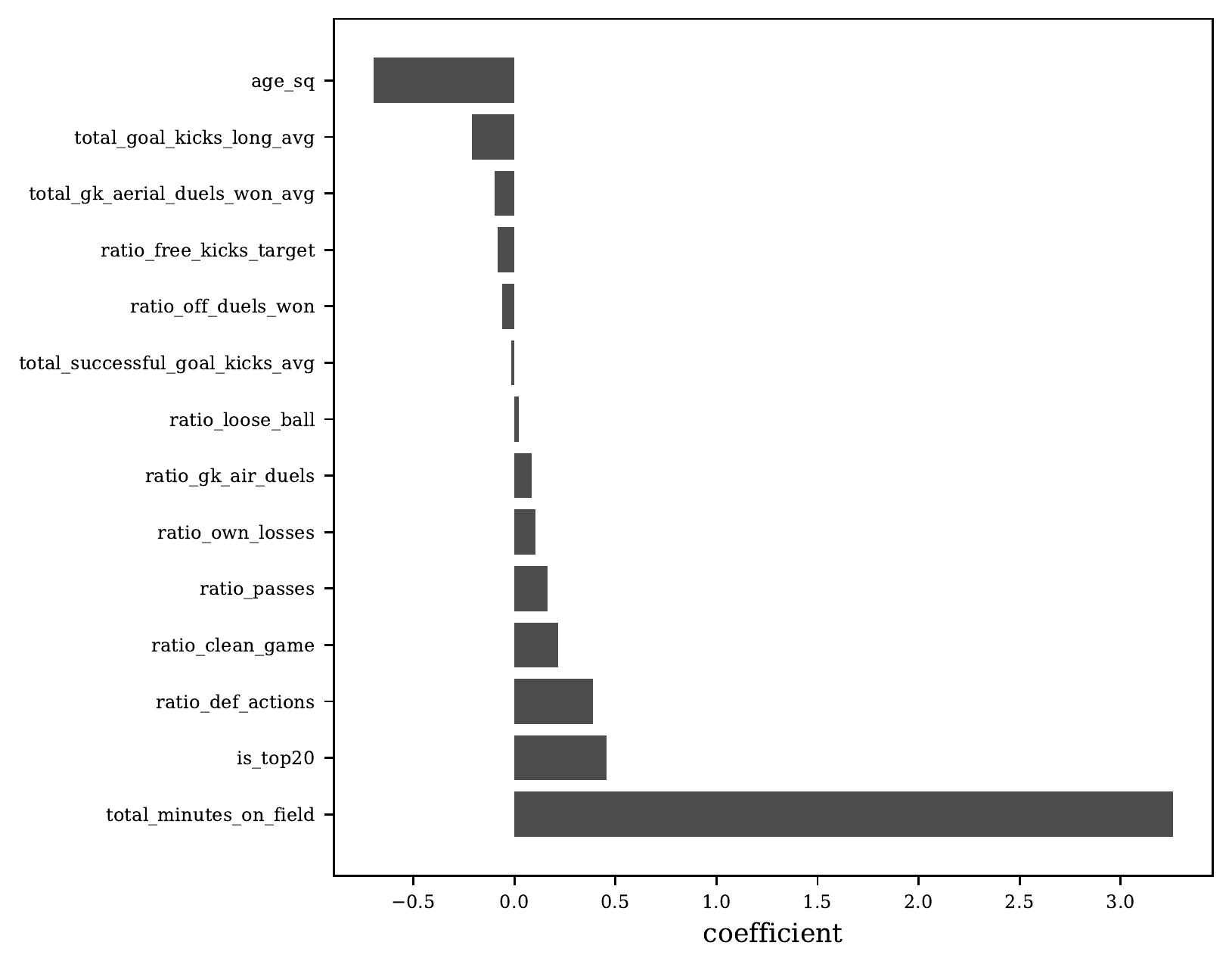}}\vspace*{-.5em}
    \end{minipage}
    \hspace{1cm}
    \begin{minipage}{0.43\textwidth}
        \centering
        \scalebox{0.4}{\includegraphics{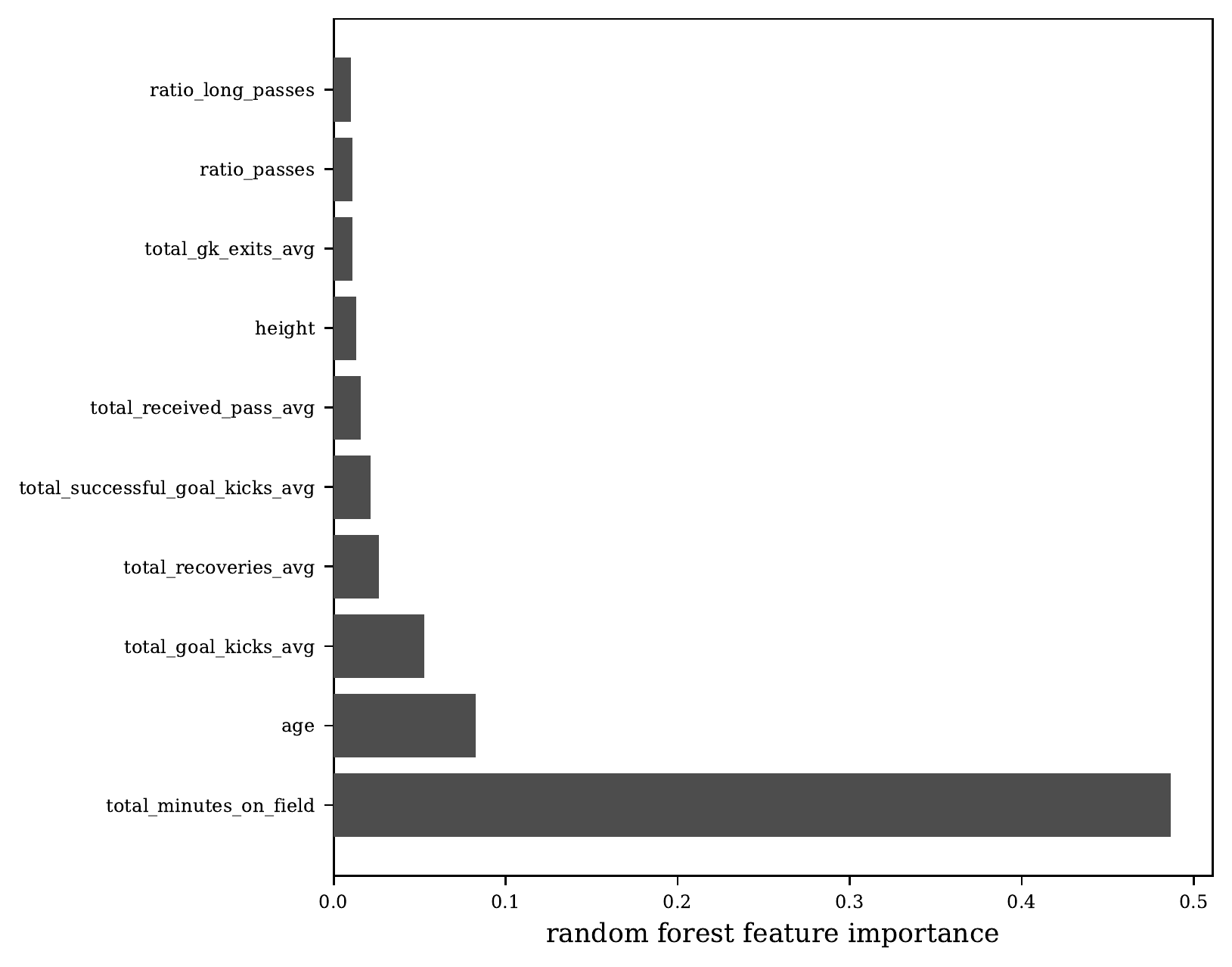}}\vspace*{-.5em}
             \end{minipage}
\caption{Important features for goalkeepers.}
\label{fig:feats_goal}
\end{figure}

Below is a list of interpretable important features for goalkeepers in both Lasso and Random Forest regressions, see Figure \ref{fig:feats_goal} for the complete list.
\begin{itemize}
    
     \item Ratio of clean sheets: keeping a high ratio of clean sheets demonstrates, among other things, the keepers' ability in saving shots. It makes sense that this feature has a strong positive coefficient.
     \item Ratio of successful goalkeeper air duels: a key element of a goalkeeper's game is to successfully intercept aerial balls using his height. This is reflected by the importance of this feature.
     \item Ratio of successful defensive actions: the goalkeeper being the last line of defense, it is not surprising that this feature is seen as a relevant one.

\end{itemize}

\subsubsection{Fullbacks}
\begin{figure}[t!]
    \centering
    \begin{minipage}{0.43\textwidth}
        \centering
        \scalebox{0.4}{\includegraphics{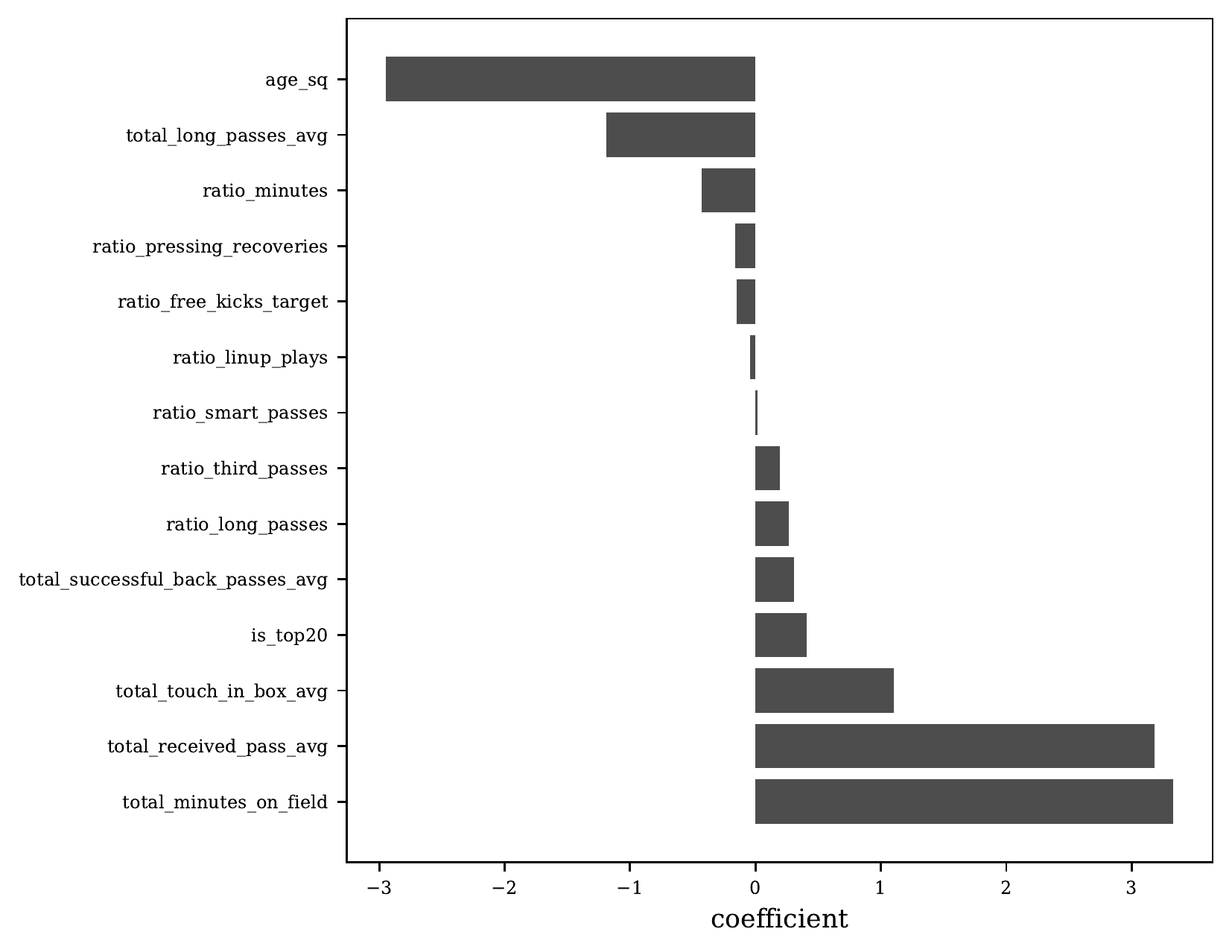}}\vspace*{-.5em}
    \end{minipage}
    \hspace{1cm}
    \begin{minipage}{0.43\textwidth}
        \centering
        \scalebox{0.4}{\includegraphics{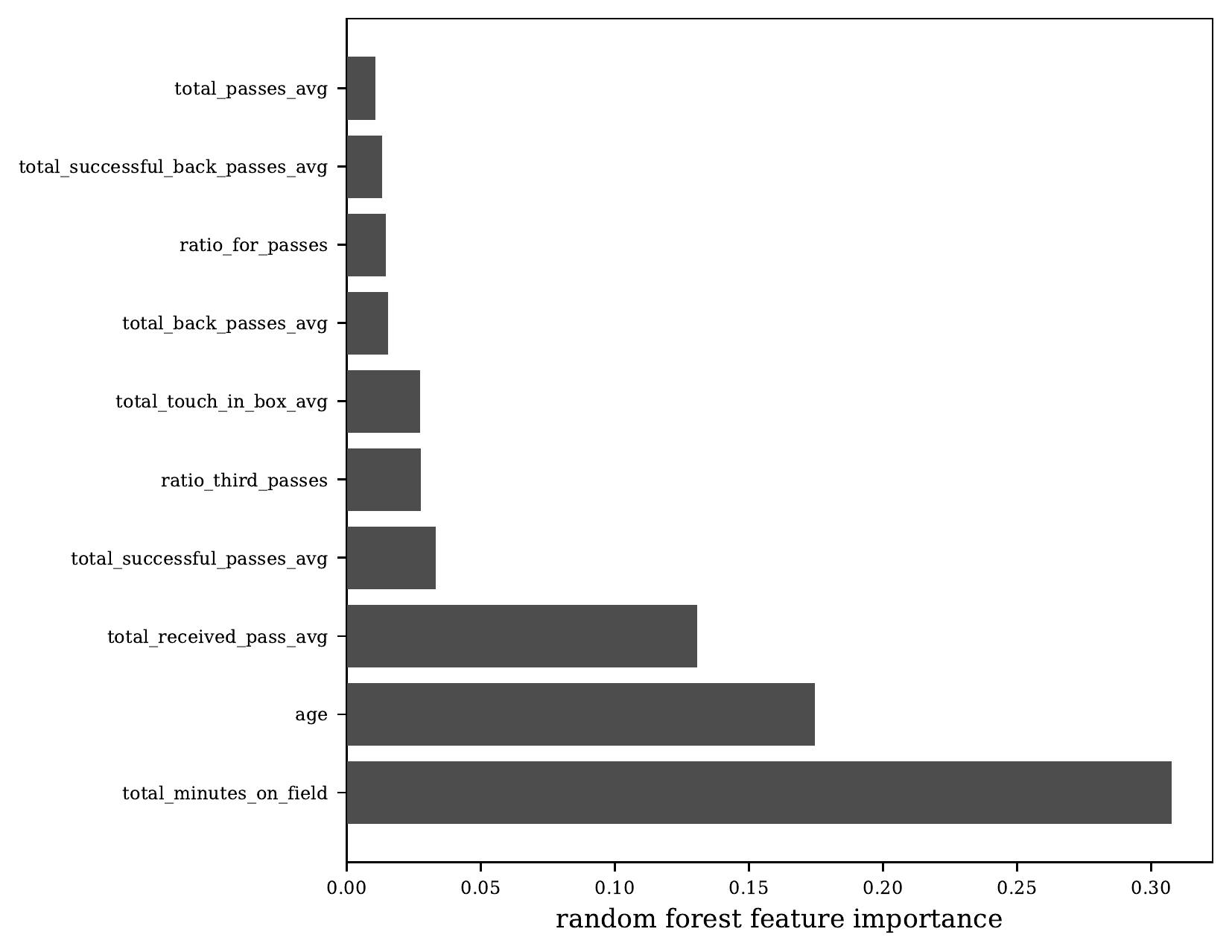}}\vspace*{-.5em}
             \end{minipage}
\caption{Important features for fullbacks.}
\label{fig:fulldef}
\end{figure}

We now interpret some of the selected features for the fullback position, see Figure \ref{fig:fulldef} for the complete list.

 \begin{itemize}

     \item Total successful passes and total successful back passes: accurate passing is a vital skill during build-up since fullbacks usually deal with the ball in dangerous areas. A valuable fullback can successfully feed the forward positions and recycle possession with back passes.
     \item Total touches in the box and ratio of successful passes to the final third: these features highlight the importance of offensive involvement for fullbacks.
    
\end{itemize}

\subsubsection{Central defenders}
\begin{figure}[t!]
    \centering
    \begin{minipage}{0.43\textwidth}
        \centering
        \scalebox{0.4}{\includegraphics{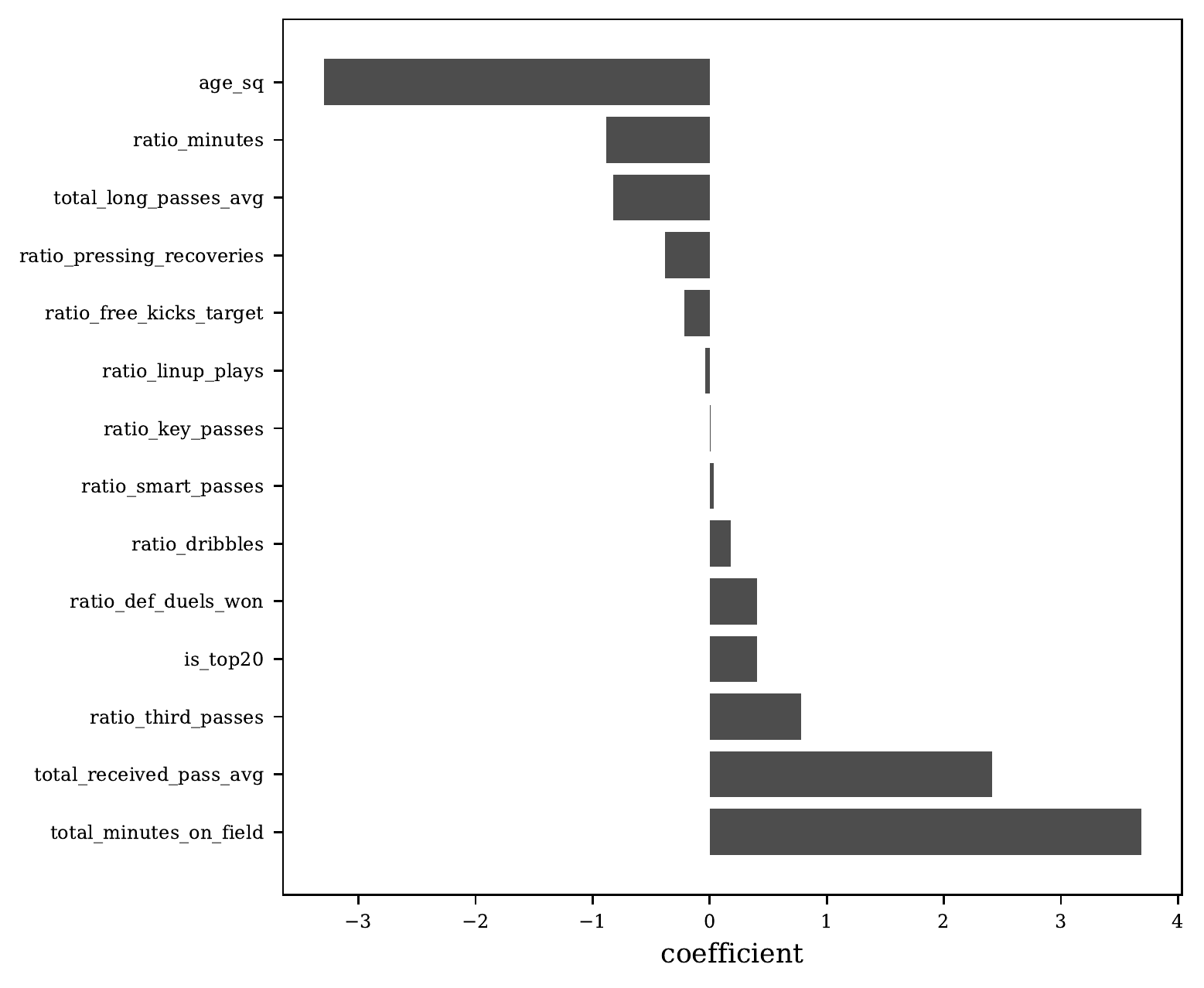}}\vspace*{-.5em}
    \end{minipage}
    \hspace{1cm}
    \begin{minipage}{0.43\textwidth}
        \centering
        \scalebox{0.4}{\includegraphics{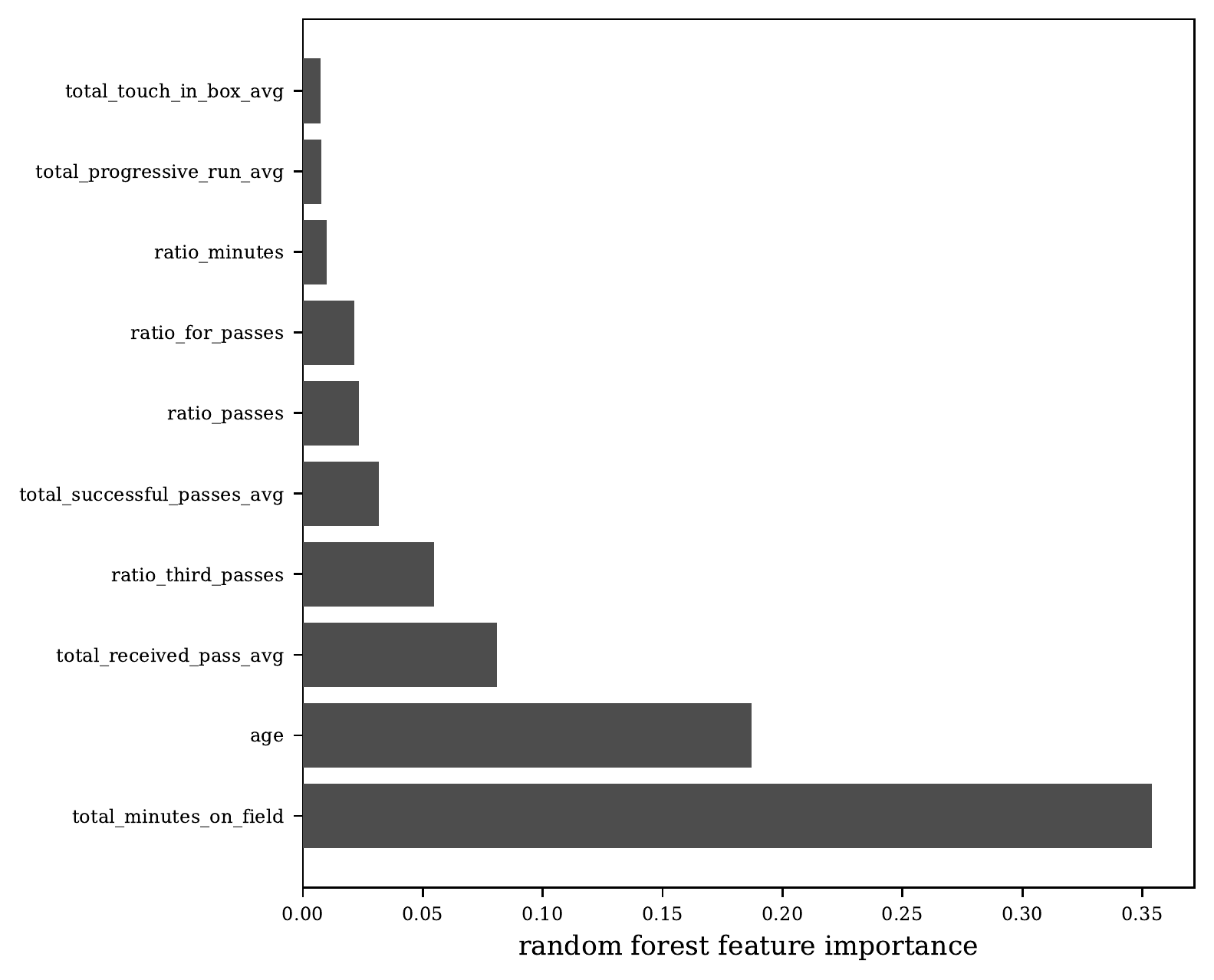}}\vspace*{-.5em}
             \end{minipage}
\caption{Important features for central defenders.}
\label{fig:cdef}
\end{figure}

We give here a list of interpretable important features for central defenders, see Figure \ref{fig:cdef} for the complete list.

 \begin{itemize}
 
        \item Ratio of defensive duels won: duels are key for center backs since a lost duel yields a possession spell for the opposition or even a scoring chance. A duel won can neutralize an opposing attack and trigger a counter-attacking opportunity.

         \item Total long passes: Surprisingly, the regression assigns a negative coefficient to this feature even if center backs generally play long. It can be explained by the fact that the tendency to play short is encouraged in modern football and the more valuable center-backs actively participate in the team build-up.
         \item Ratio of successful passes to the final third: it is very unlikely that a center back gives a key pass. However, the ability to deliver balls in the final third significantly helps build up and enables the team to create chances. It is not surprising that this feature is selected.
\end{itemize}
 
\subsubsection{Defensive midfielders}
\begin{figure}[t!]
    \centering
    \begin{minipage}{0.43\textwidth}
        \centering
        \scalebox{0.4}{\includegraphics{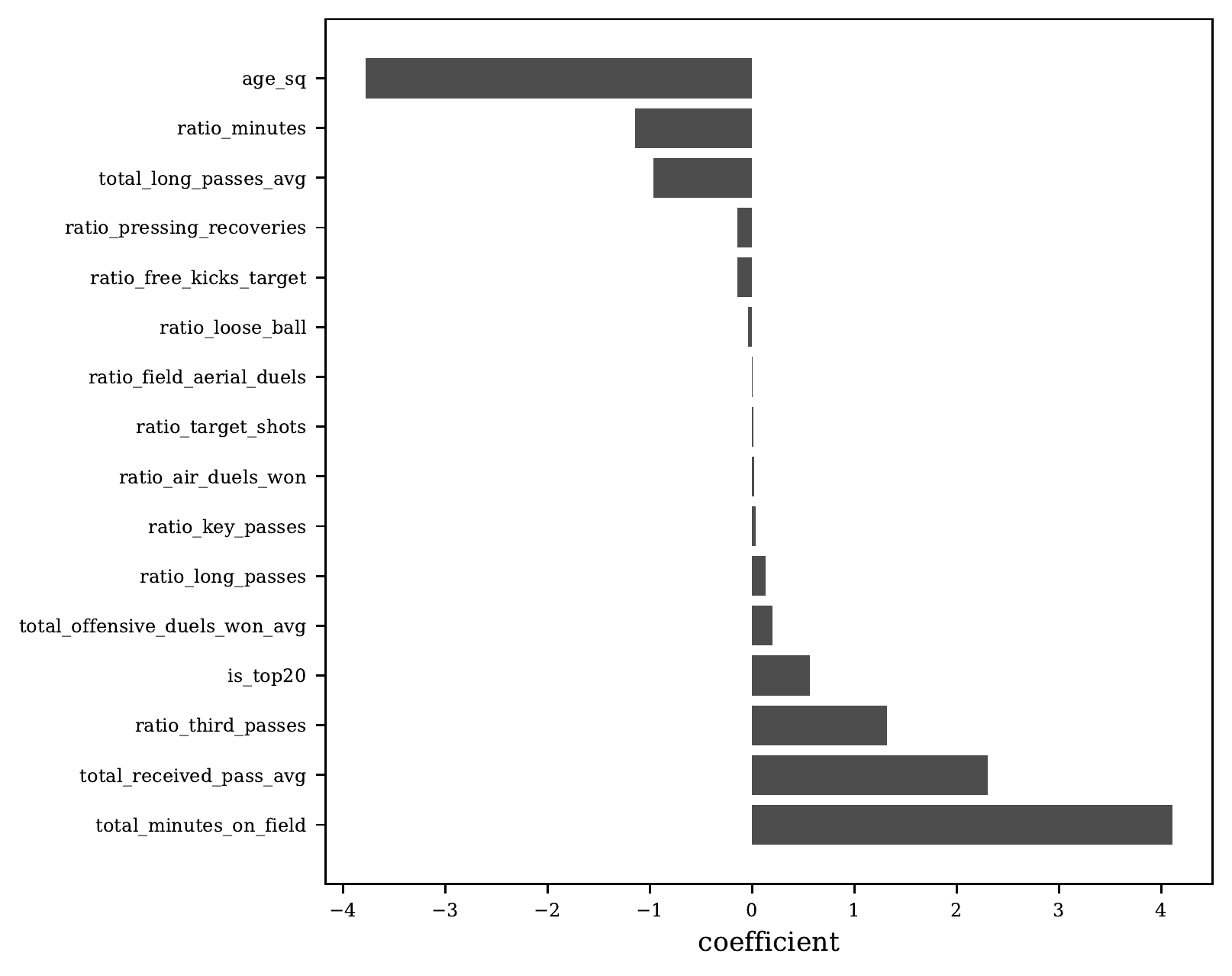}}\vspace*{-.5em}
    \end{minipage}
    \hspace{1cm}
    \begin{minipage}{0.43\textwidth}
        \centering
        \scalebox{0.4}{\includegraphics{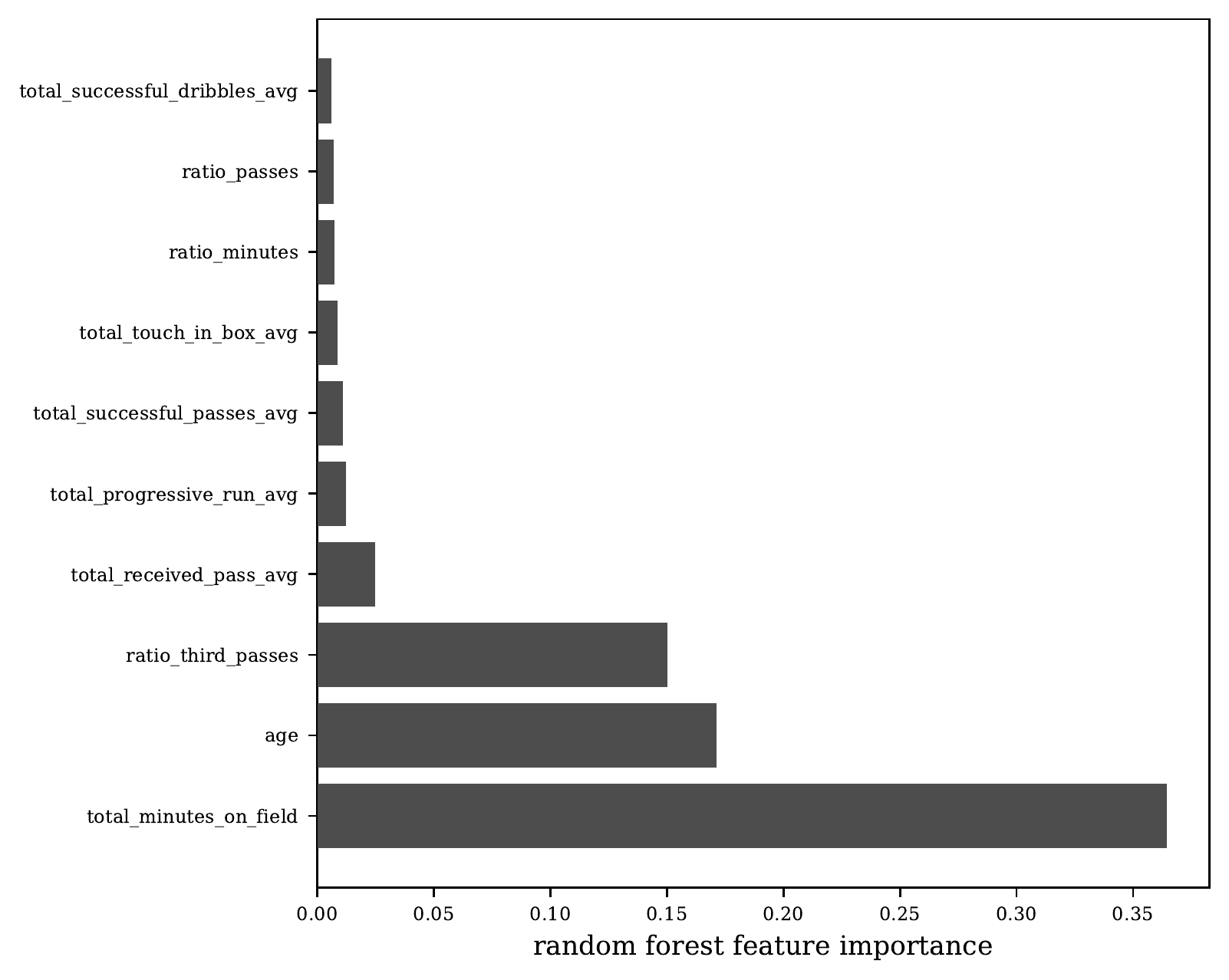}}\vspace*{-.5em}
             \end{minipage}
\caption{Important features for defensive midfielders.}
\label{fig:defmid}
\end{figure}

We discuss now the prominent features for defensive midfielders, see Figure \ref{fig:defmid} for the complete list.
 \begin{itemize}
     \item Ratio of successful passes to the final third: the defensive midfielders have to keep the team structured together in defense and be able to actively distribute the ball forward. Their role is to find wingers and attacking midfielders that can then create chances. The positive coefficient of this feature reflects the importance of progressive play in this position.
     \item Total offensive duels won: defensive midfielders are the trigger for pressing and also a crucial part of winning second balls to avoid counters. Unsurprisingly, this feature stands out.
     \item Total long passes and ratio of successful long passes: attempting long balls again has a negative impact on the player's value. This time, we also observe a small positive coefficient for the ratio of successful long passes. This highlights the importance of long-range ability for this position. For example, Xabi Alonso can deliver very precise long balls but only relies on them when he spots interesting opportunities. 
\end{itemize}
 
\subsubsection{Central midfielders}
We now turn to the case of central midfielders. Figure \ref{fig:mid} shows a list of variables selected by the Lasso and Random Forest regressions. Below is the interpretation of some of those features:

 \begin{figure}[t!]
    \centering
    \begin{minipage}{0.43\textwidth}
        \centering
        \scalebox{0.4}{\includegraphics{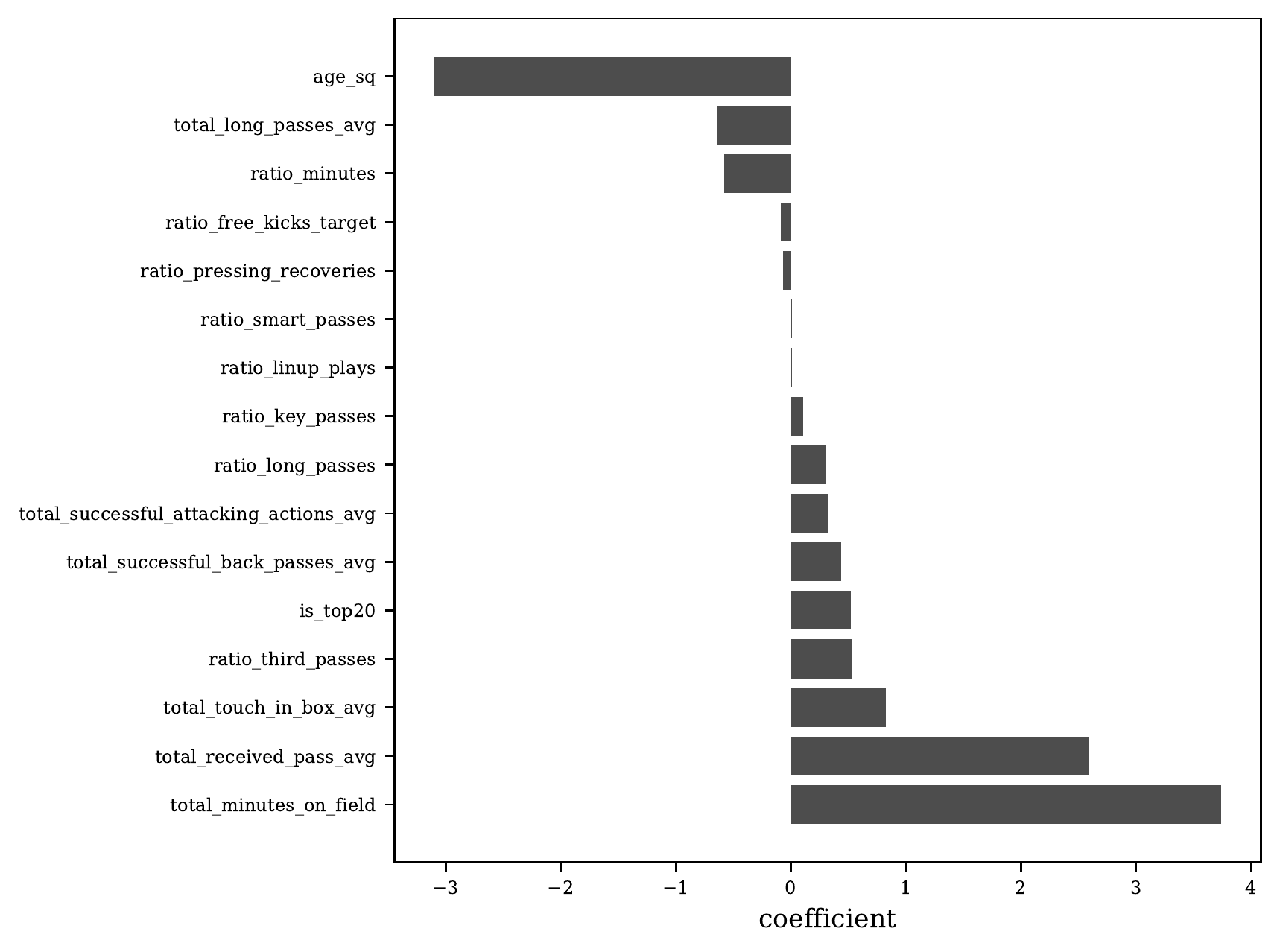}}\vspace*{-.5em}
    \end{minipage}
    \hspace{1cm}
    \begin{minipage}{0.43\textwidth}
        \centering
        \scalebox{0.4}{\includegraphics{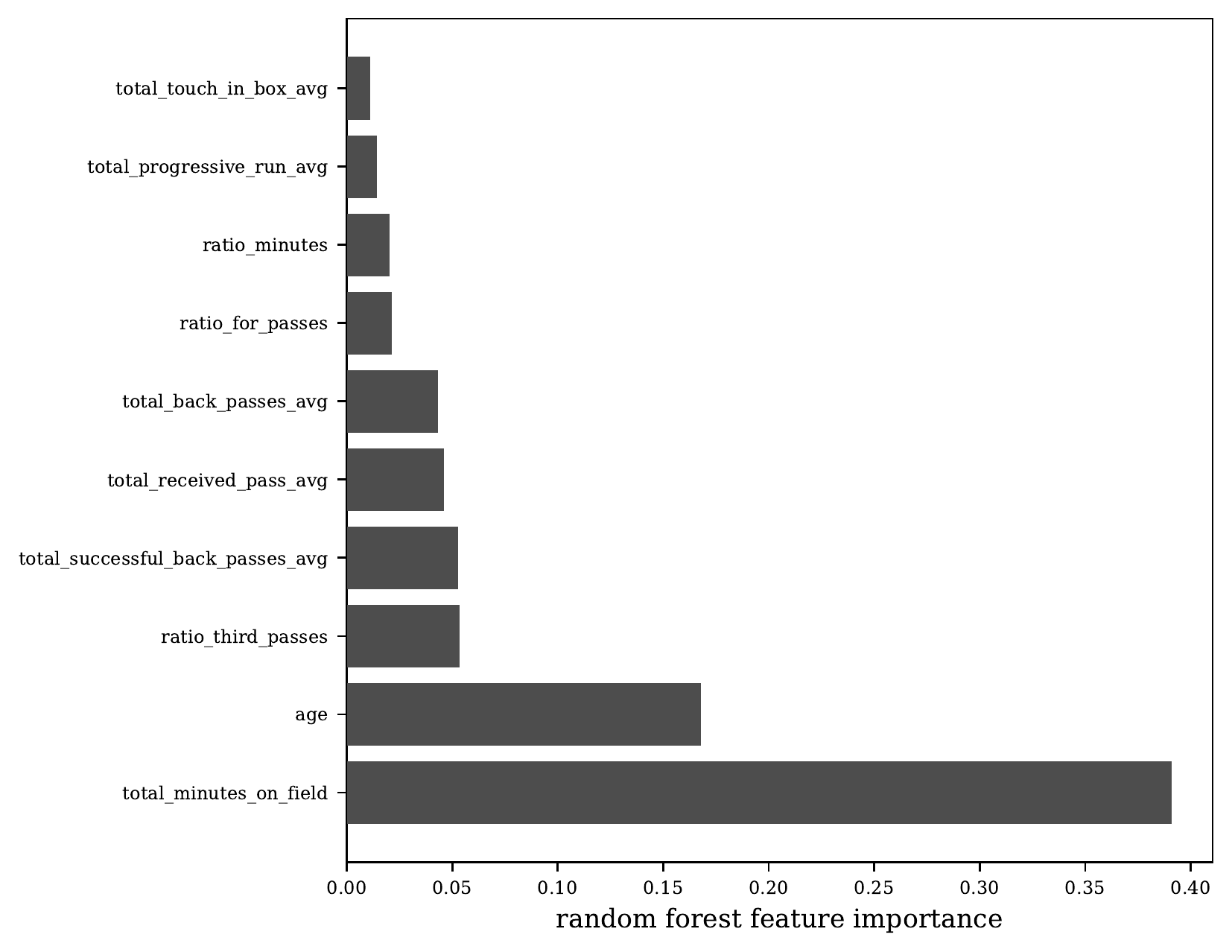}}\vspace*{-.5em}
             \end{minipage}
\caption{Important features for central midfielders.}
\label{fig:mid}
\end{figure}

\begin{itemize}
    \item Total received passes: a central midfielder's involvement is the most important aspect of his game and it only comes second to minutes played in our estimator.
    
    \item Total successful attacking actions and total touches in the opponent box: central midfielders can have various game styles. Usually, the most valuable players are the ones who have good vision going forward.
    
    \item Total long passes and ratio of successful long passes: the tendency to play short passes and keep possession is important in a midfielder's valuation. Again, the rate of success in long balls can add some value to the player.
\end{itemize}

\subsubsection{Attacking midfielders}
\begin{figure}[t!]
    \centering
    \begin{minipage}{0.43\textwidth}
        \centering
        \scalebox{0.4}{\includegraphics{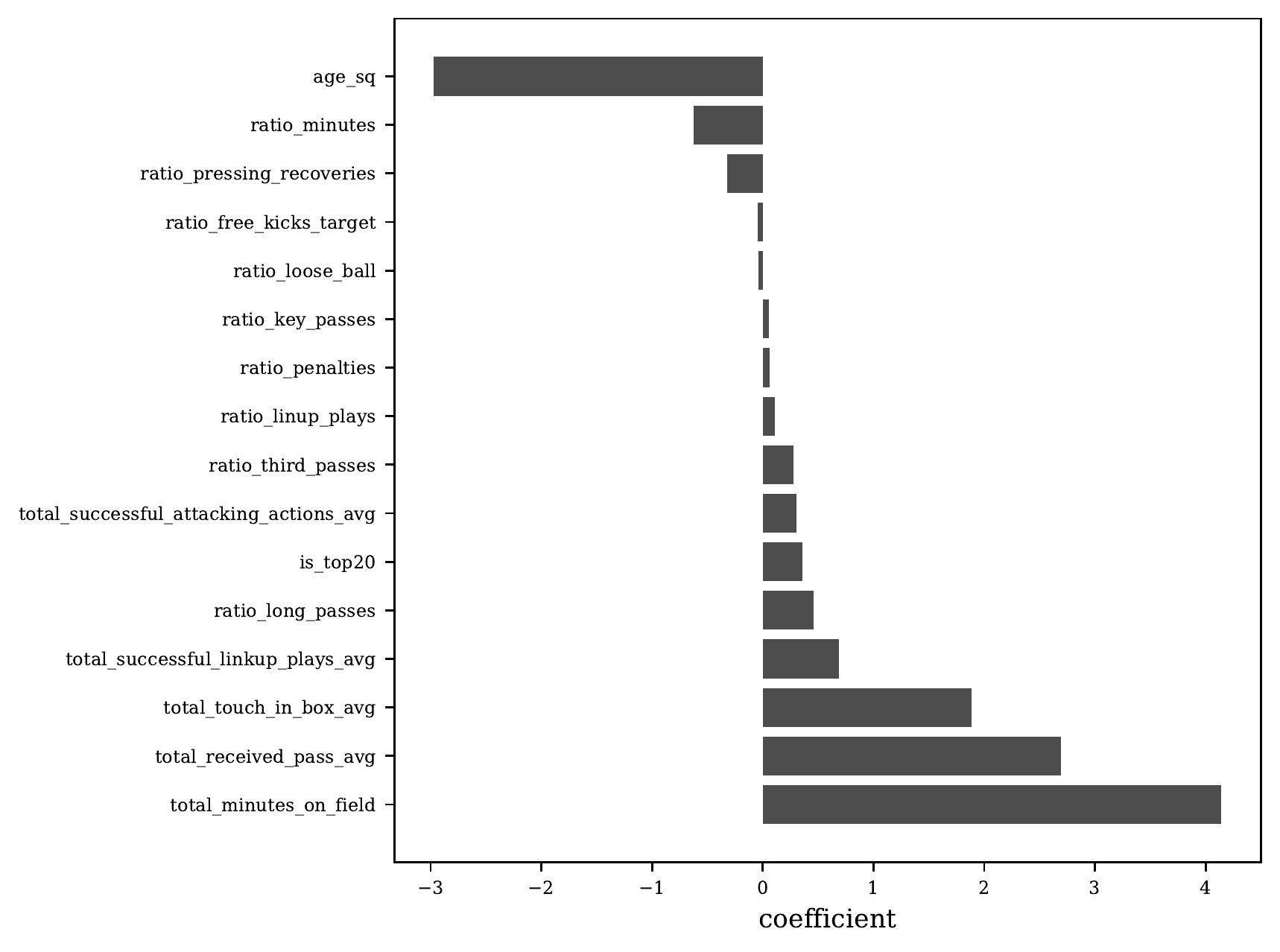}}\vspace*{-.5em}
    \end{minipage}
    \hspace{1cm}
    \begin{minipage}{0.43\textwidth}
        \centering
        \scalebox{0.4}{\includegraphics{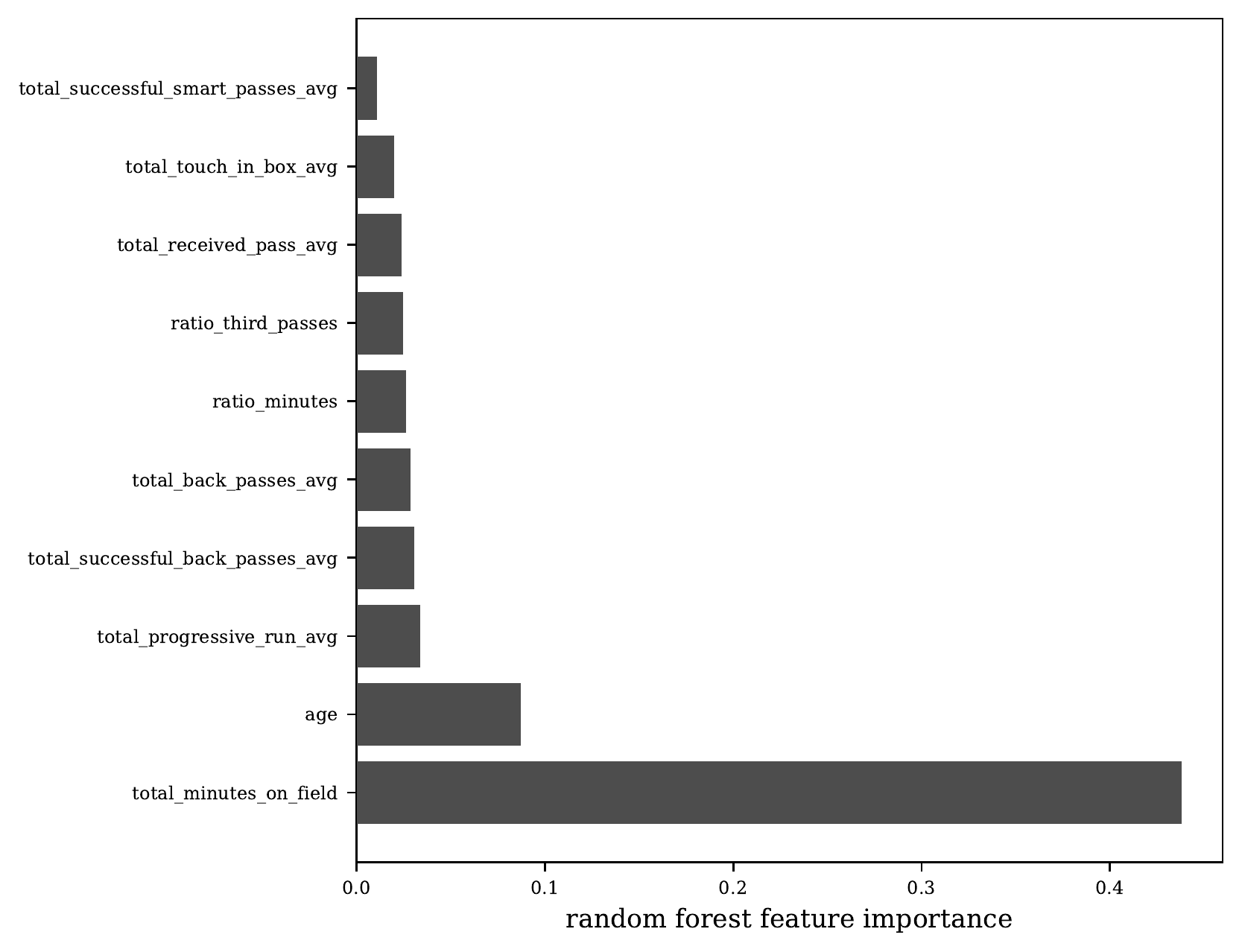}}\vspace*{-.5em}
             \end{minipage}
\caption{Important features for attacking midfielders.}
\label{fig:attmid}
\end{figure}

We now look at the prominent predictors for attacking midfielders, see Figure \ref{fig:attmid} for the complete list.

 \begin{itemize}
     \item Total touches in the opponent box: this feature ranks very high for attacking midfielders and it shows the importance of the offensive aspect of their game. The closer they are to the box, the higher the probability of scoring goals.
   
     \item Total successful linkup plays, successful attacking actions, and total progressive runs: the presence of these features highlights the play-making aspect of attacking midfielders. In particular, their task is to combine in small spaces and advance the ball to dangerous areas.
 \end{itemize}

\subsubsection{Wingers}
\begin{figure}[t!]
    \centering
    \begin{minipage}{0.43\textwidth}
        \centering
        \scalebox{0.4}{\includegraphics{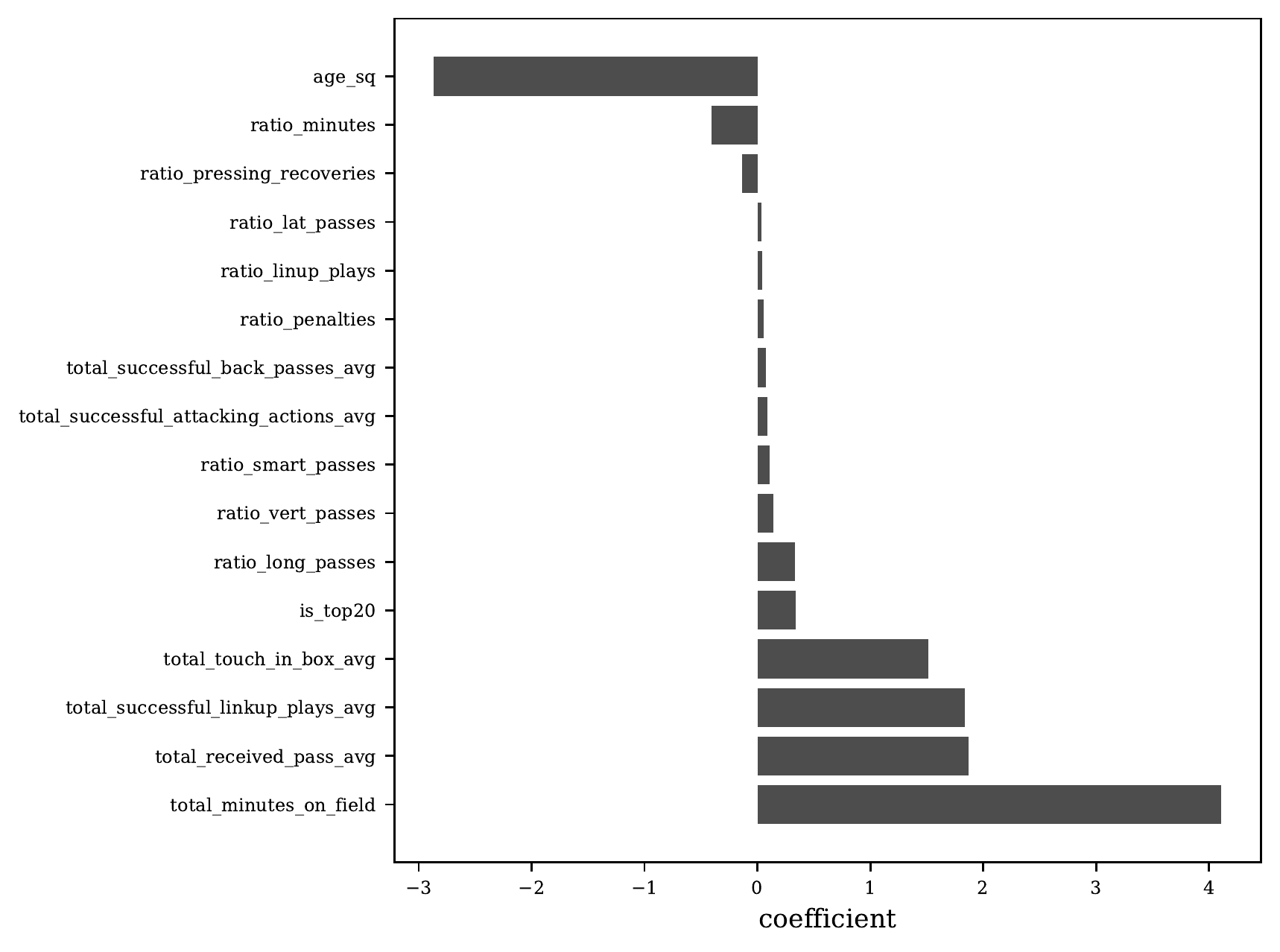}}\vspace*{-.5em}
    \end{minipage}
    \hspace{1cm}
    \begin{minipage}{0.43\textwidth}
        \centering
        \scalebox{0.4}{\includegraphics{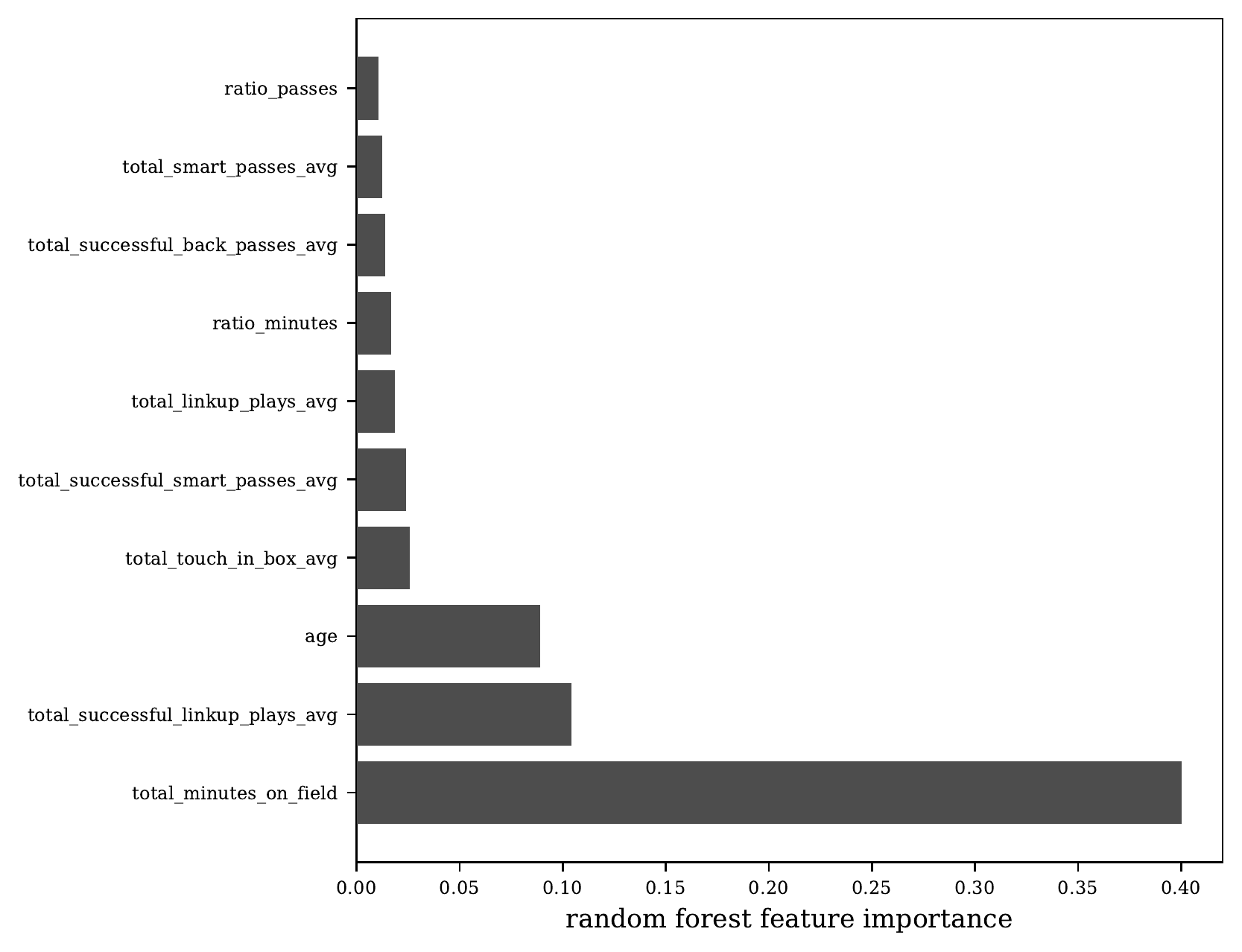}}\vspace*{-.5em}
             \end{minipage}
\caption{Important features for wingers.}
\label{fig:wing}
\end{figure}

Here, we provide a list of important features for wingers along with their interpretation, see Figure \ref{fig:wing} for the complete list.

 \begin{itemize}
    \item Total received passes: wingers are positioned wide on the pitch and good positioning to receive passes is a crucial part of their role to spread opposition and help build up.
    \item Total touches in the opponent box: the more valuable wingers get involved in the final third and try to score goals. It makes sense that this feature has a positive coefficient. 
    \item Total successful linkup plays: it is important for wingers to combine with the attacking midfielders and center-forward to make key passes or score goals. It is not surprising that this feature is selected for this position.
 \end{itemize}
 
\subsubsection{Forwards}
\begin{figure}[t!]
    \centering
    \begin{minipage}{0.43\textwidth}
        \centering
        \scalebox{0.4}{\includegraphics{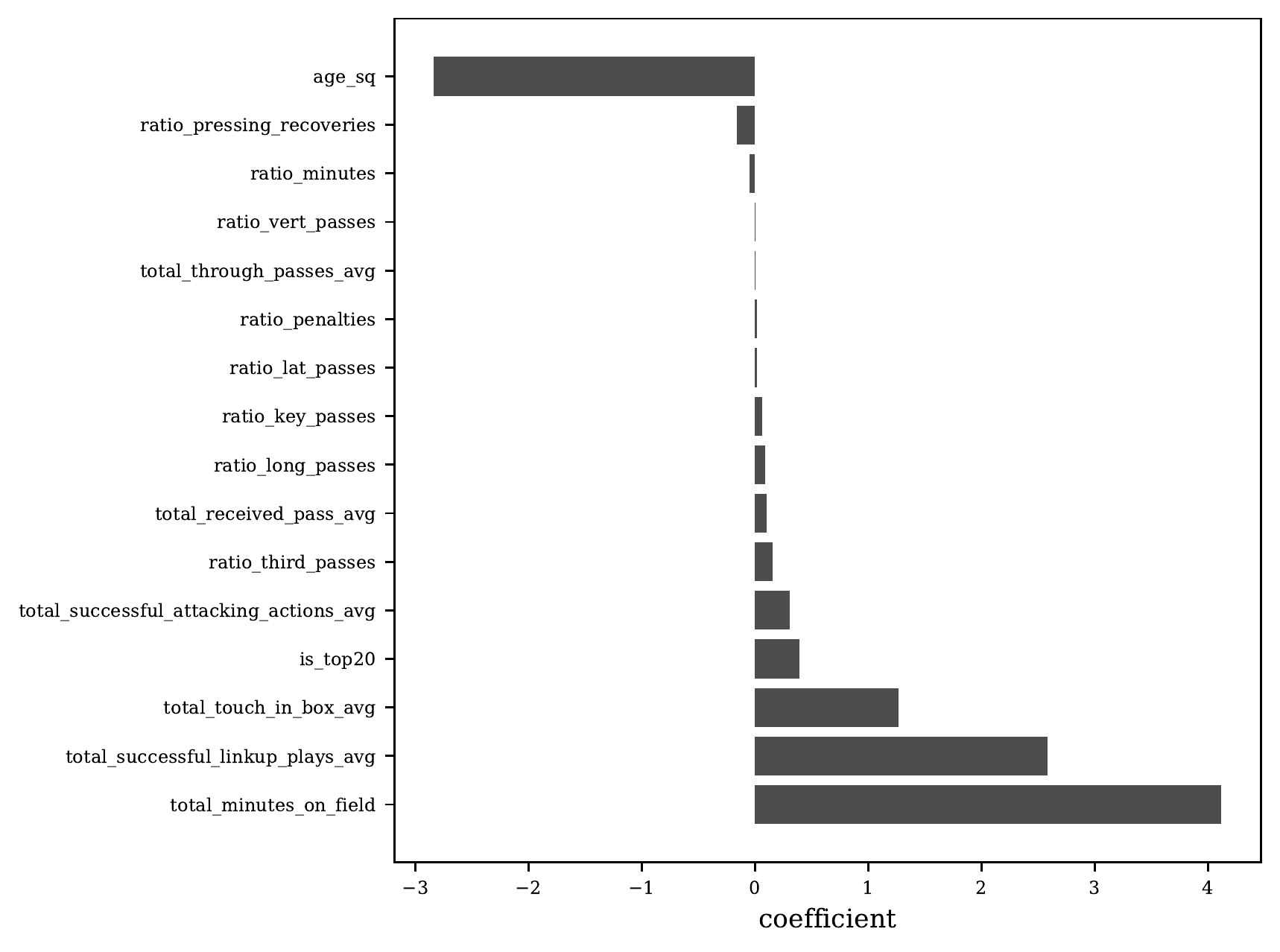}}\vspace*{-.5em}
    \end{minipage}
    \hspace{1cm}
    \begin{minipage}{0.43\textwidth}
        \centering
        \scalebox{0.4}{\includegraphics{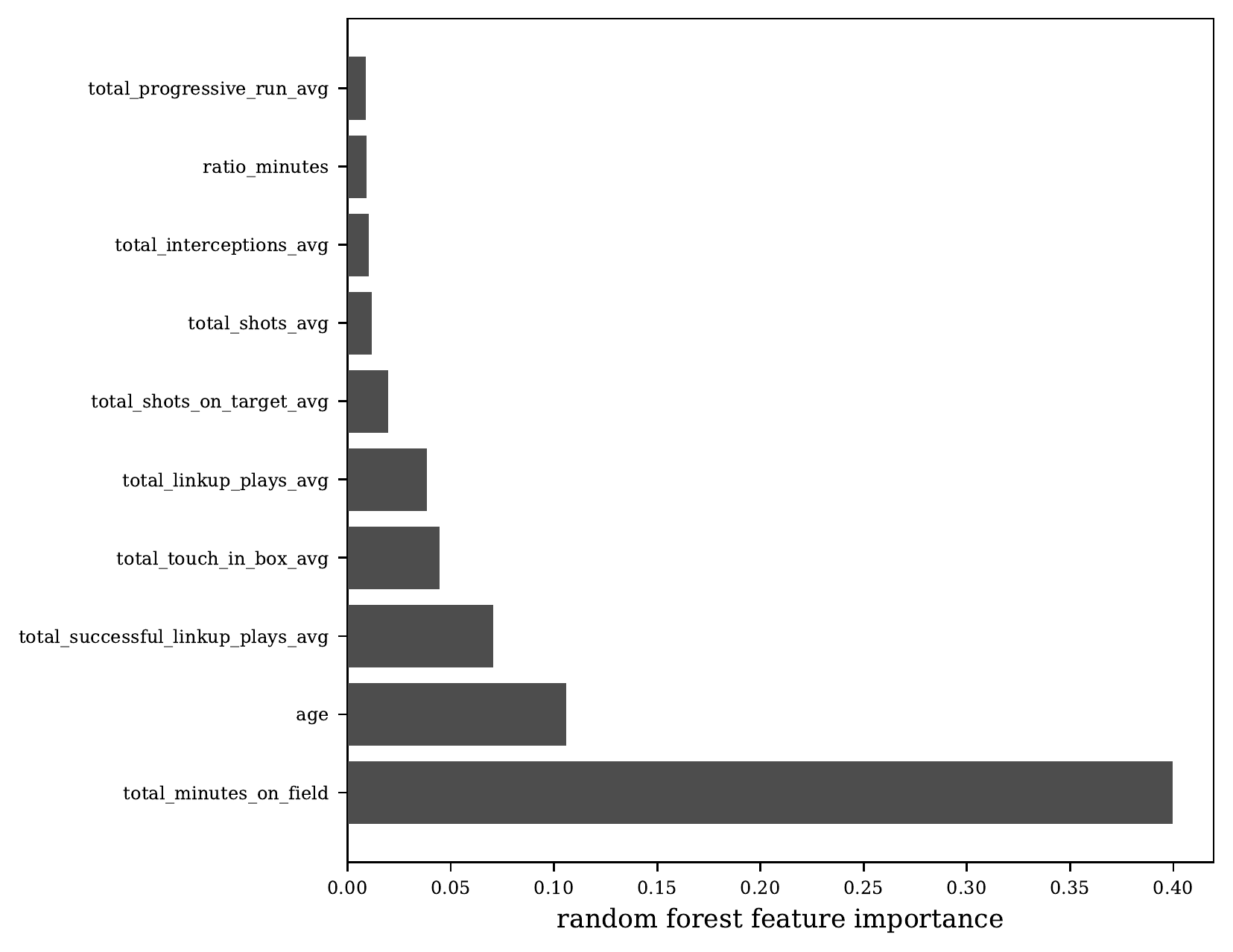}}\vspace*{-.5em}
             \end{minipage}
\caption{Important features for forwards.}
\label{fig:att}
\end{figure}

Below is a list of interpretable important features for forwards in both Lasso and Random Forest regressions, see Figure \ref{fig:att} for the complete list.
 \begin{itemize}
     \item Total successful linkup plays: this feature reflects the striker's ability to combine in small spaces and help advance the ball to more dangerous areas. 
     \item Total touches in the opponent box: a touch in the box is an opportunity for scoring and indicates that the center-forward is well-positioned on the field and makes himself available in dangerous zones of the opposition. It makes sense that this feature is considered important.
     \item Total successful attacking actions and total shots on target: Both these features are important for strikers as they reflect the quality of their offensive output and finishing.
 \end{itemize}

 Surprisingly, the number of goals scored is not selected by our regression. It can be explained by the presence of several other features that are very correlated to goals. In particular, the features cited above are significantly connected to scoring goals but provide extra information on the offensive contribution.
\paragraph{Remark}
We observe that some features have a small negative coefficient in our Lasso regression despite being positively correlated to the target, or intuitively having positive causation. This is due to the fact our features are very correlated and some features are sometimes used as a non-linear correction to the effect of another feature. One example is the negative coefficient of $ratio\_minutes =\frac{total\_minutes}{total\_matches}$ while $total\_minutes$ has a high positive coefficient. The regressor is using it to correct the dependency on $total\_minutes$ and account for a non-linear positive dependency on $total\_matches$.

\section{Predicting value for young players}
\label{sec:youngplayers}
A particular application of our methodology is to quantify young players' potential. To capture features specifically relevant for young players, we fit a new Lasso regression on young players to predict their future market value. Given the fact that the Wyscout dataset mainly covers the window between 2018 and 2022 and that a young player's valuation can significantly change in a short span of time, we forecast the future market value in a one year horizon using performance statistics from the previous year. Thus, we only keep in the training and validation datasets the players for whom we had data when they were under 21, with $t$ being the time of their last available game before turning 22. Another condition is to have a Transfermarkt value available at time $t+1\ \textup{year}$. Furthermore, as a lot of young players play in second or third division leagues, we only keep statistics from games played in first division leagues. Finally, since the resulting dataset has a small number of datapoints (951), it is necessary to perform a fit on all positions simultaneously by adding the position as a feature using one hot encoding.

The cross-validation score stays in the same range with the important features being \textit{total minutes on field}, \textit{average league value}, \textit{age}, \textit{total touches in box} and \textit{total successful passes} average. To test our approach, we forecast the market value of the nominees to the Golden Boy 2022 award and compare the resulting value-based ranking with the choice of the jury. The Golden Boy is awarded  every year to the best talent under the age of 21 based on the votes of past Ballon d'Or winners. 

Table \ref{tab:goldenboys} shows the resulting top thirteen based on our predicted future value, the prediction is performed on the sixty nominees available in our dataset, see Table \ref{tab:fullranking} for the complete list of nominees. Note that Pedri is ranked second and he's the winner of the 2021 edition. Table \ref{tab:truegoldenboys} shows the resulting top ten voted by the jury. Nine out of the realized top ten are included in our predicted top thirteen. Traditionally, previous winners of the Golden Boy are not in contention for the prize the following year. This explains the absence of Pedri in the jury ranking despite his proven quality and our approach placing him second.

\begin{table}[t!]
\begin{center}
    \begin{tabular}{c c c}
    \toprule
         \textbf{Predicted Rank} &\textbf{Name} & \textbf{Club}\\ 
         \midrule
         1 & Bukayo Saka & Arsenal \\

        2 & Pedri & FC Barcelona \\ 
        3 & Jude Bellingham & Borussia Dortmund  \\ 
        4 & Josko Gvardiol & RB Leipzig  \\ 
        5 & Eduardo Camavinga & Real Madrid  \\ 
        6 & Gavi & FC Barcelona  \\ 
        7 & Jamal Musiala & Bayern Munich\\ 
        8 & Arnaud Kalimuendo & Stade Rennais  \\ 
        9 & Nuno Mendes & PSG  \\ 
        10 & Kamaldeen Sulemana & Stade Rennais   \\
        11 & Florian Wirtz & Bayer Leverkusen\\
        12 & Nico Williams & Athletic Club  \\ 
        13 & Ryan Gravenberch  & Bayern Munich\\ 
          \bottomrule
    \end{tabular}
    \end{center}
    \vspace{0.4em}
    \caption{Top thirteen based on the predicted value in one year.}\label{tab:goldenboys}
\end{table}

\begin{table}[t!]
\begin{center}
    \begin{tabular}{c c c}
    \toprule
         \textbf{Rank} &\textbf{Name} & \textbf{Club}\\ 
         \midrule
         1 & Gavi & FC Barcelona \\ 
        2 & Eduardo Camavinga & Real Madrid \\ 
        3 & Jamal Musiala & Bayern Munich\\ 
        4 & Jude Bellingham  & Borussia Dortmund \\ 
        5 & Nuno Mendes & PSG \\ 
        6 & Josko Gvardiol & RB Leipzig   \\ 
        7 & Ryan Gravenberch & Bayern Munich\\ 
        8 & Bukayo Saka & Arsenal\\ 
        9 & Karim Adeyemi & Borussia Dortmund  \\ 
        10 & Florian Wirtz & Bayer Leverkusen \\ 
          \bottomrule
    \end{tabular}
    \end{center}
    \vspace{0.4em}
    \caption{Top ten based on Golden Boy jury votes.}\label{tab:truegoldenboys}
\end{table}  

\begin{figure}[t!]
    \centering
    \begin{minipage}{0.95\textwidth}
        \centering
        \scalebox{0.80}{\includegraphics{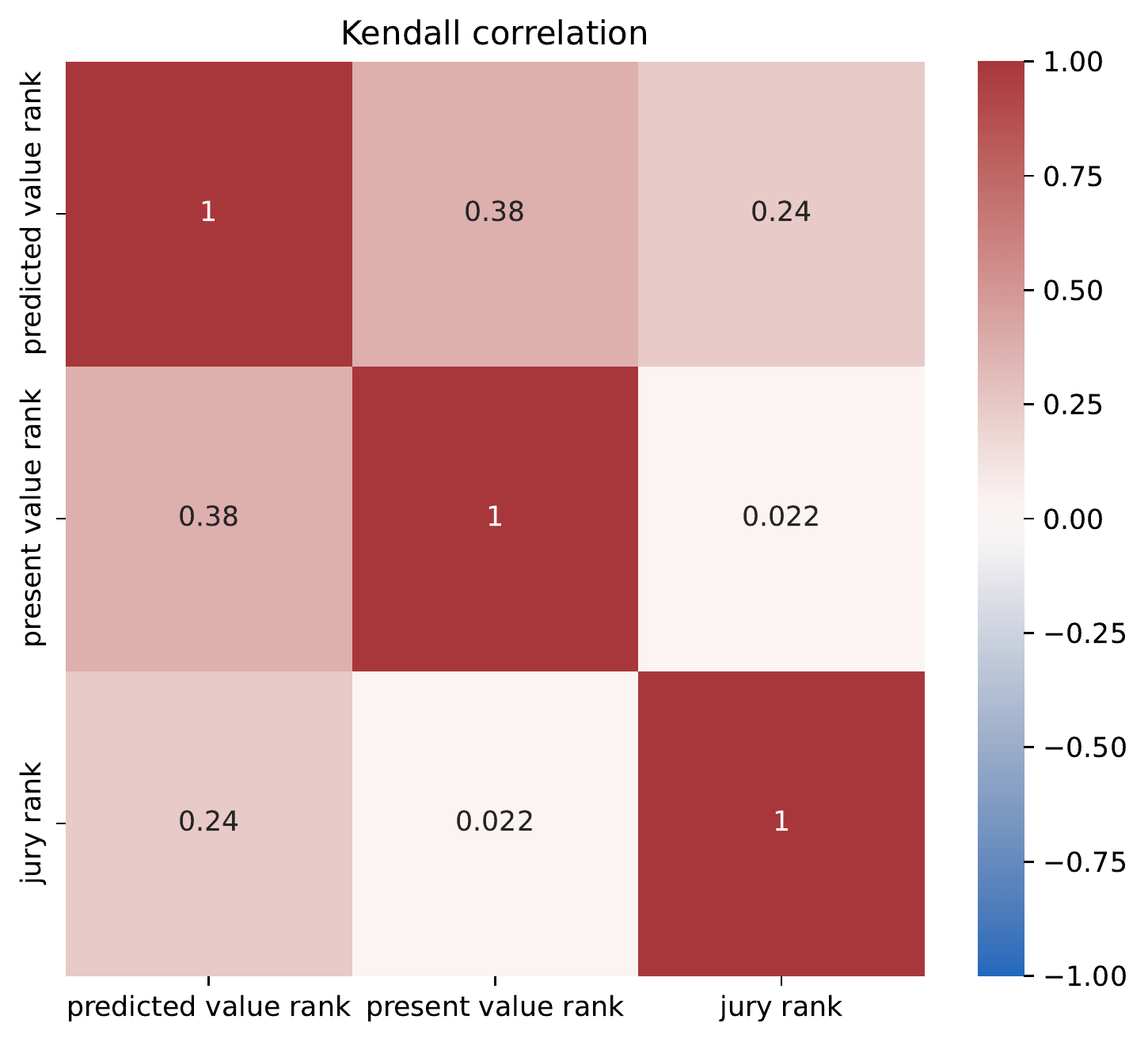}}\vspace*{-.8em}
        \captionof{figure}{Correlation between the jury ranking and rankings based on predicted value and current value.}  \label{fig:correlation} 
    \end{minipage}

\end{figure}

Figure \ref{fig:correlation} shows the Kendall correlations between the jury ranking and the rankings based on our predicted market value and the present one. The correlations with the jury ranking are computed on the released top ten players only while the correlation between value-based rankings (predicted and present market value) is computed using all sixty nominees. We observe that our approach has a significant correlation with the present value ranking. This is not surprising because we forecast the market valuation in one year which is obviously connected to the present Transfermakt one. Moreover, our predicted order shows a significantly higher correlation to the jury ranking compared to the ranking based on the current TransferMarkt value. This shows that our methodology  does eliminate noise from the pricing of players related to external factors such as fan hype or unreasonable club demand, and isolates the value directly related to player performance and in-game metrics.

%% file: sections/conclusion.tex
\section{Conclusion and future work}
Using in-game performance data and player characteristics, we develop regressions to forecast the future market value of football players and determine the relevant features involved in this prediction according to the position. The results are very promising as we achieve an explained variance as high as 60\%. We uncover obvious prominent variables for all positions but also some subtle dependencies. One should keep in mind this is not a procedure to predict the true value in the future because we ignore a lot of important factors, but a quantitative process to identify prospective talent and compare players based on quality and performance. As an application, we show our methodology can successfully rank young players with respect to their quality. Future work will incorporate momentum indicators and external factors such as popularity to improve forecast accuracy. The effect of the league on the pricing of football players is to be investigated further as well. In particular, different characteristics are valued in each league depending on the football culture of the country. Such differences can significantly affect a player's performance and market value after a transfer.

%% file: sections/annex.tex
\newpage
\section{Annex}

In the annex, we display a list of the features provided in the Wyscout dataset, a list of the engineered ratio features as well as our predicted ranking of the Golden boy nominees.

\begin{table}[ht]
        \centering
        \resizebox{\linewidth}{!}{
        \begin{tabular}{ll}
        \toprule
        {} & \textbf{Explanation} \\
        \midrule
        player\_id\_transfermarkt  &      The player's id in the TransferMarkt dataset       \\
        match\_date        &     The date of the game    \\
        total\_matches    &     1 if the player played, 0 otherwise   \\
        total\_minutes\_on\_field  & Number of minutes played during the game by the player \\
        total\_goals    & Number of goals scored during the game    \\
        total\_assists   & Number of assists completed during the game   \\
        total\_shots   & Number of shots attempted during the game  \\
        total\_head\_shots    & Number of headers attempted     \\
        total\_yellow\_cards     &  Number of yellow cards given to the player during the game    \\
        total\_red\_cards   & Number of red cards given to the player during the game   \\
        total\_direct\_red\_cards   &  Number of direct red cards given to the player (not after two yellows) \\
        total\_penalties   &  Number of penalties attempted  \\
        total\_linkup\_plays  &   Number of linkup plays by the player  \\
        total\_duels   & Number of duels taken by the player \\
        total\_duels\_won  & Number of duels won by the player   \\
        total\_defensive\_duels & Number of defensive duels attempted by the player \\
        total\_defensive\_duels\_won  & Number of defensive duels won by the player  \\
        total\_offensive\_duels & Number of offensive duels attempted by the player  \\
        total\_offensive\_duels\_won   & Number of offensive duels won by the player  \\
        total\_aerial\_duels  & Number of aerial duels attempted by the player \\
        total\_aerial\_duels\_won & Number of aerial duels won by the player  \\
        total\_fouls  &  Number of fouls done by the played \\
        total\_passes  & Number of passes attempted by the player\\
        total\_successful\_passes  & Number of passes completed by the player \\
        total\_smart\_passes   &  Number of smart passes attempted by the player \\
        total\_successful\_smart\_passes   &  Number of smart passes completed by the player   \\
        total\_passes\_to\_final\_third  &  Number of passes to the final third attempted by the player      \\
        total\_successful\_passes\_to\_final\_third   &  Number of passes to the final third completed by the player  \\
        total\_crosses  & Number of crosses attempted by the player   \\
        total\_successful\_crosses & Number of crosses completed by the player  \\
        total\_forward\_passes  & Number of forward passes attempted by the player  \\
        total\_successful\_forward\_passes   & Number of forward passes completed by the player  \\
        total\_back\_passes  & Number of back passes attempted by the player  \\
        total\_successful\_back\_passes  &  Number of successful back passes completed by the player  \\
        total\_through\_passes  & Number of through passes attempted by the player  \\
        total\_successful\_through\_passes  & Number of through passes completed by the player \\
        total\_key\_passes & Number of key passes attempted by the player \\
        total\_successful\_key\_passes & Number key passes completed by the player \\
        total\_vertical\_passes & Number of vertical passes attempted by the player \\
        total\_successful\_vertical\_passes & Number of vertical passes completed by the player \\
        total\_long\_passes  & Number of long passes attempted by the player   \\
        total\_successful\_long\_passes & Number of long passes completed by the player  \\
        \bottomrule
        \end{tabular}}
        \vspace{0.5em}
        \caption{Wyscout list of game statistics (part I)}
        \label{tab:wyscoutglos1}
    \end{table}

    \begin{table}[ht]
        \centering
        \resizebox{\linewidth}{!}{
        \begin{tabular}{ll}
        \toprule
        {} & \textbf{Explanation} \\
        \midrule
        
        total\_dribbles   & Number of dribbles attempted by the player \\
        total\_successful\_dribbles &  Number of dribbles completed by the player   \\
        total\_interceptions   & Number of interceptions completed by the player\\
        total\_defensive\_actions  & Number of defensive actions attempted by the player  \\
        total\_successful\_defensive\_actions  & Number of defensive actions completed by the player  \\
        total\_attacking\_actions  & Number of attacking actions attempted by the player   \\
        total\_successful\_attacking\_actions       & Number of attacking actions completed by the player   \\
        total\_free\_kicks  & Number of free-kicks attempted by the player  \\
        total\_free\_kicks\_on\_target & Number of on-target free-kick shots completed by the player   \\
        total\_direct\_free\_kicks    & Number of direct free-kicks attempted by the player    \\
        total\_direct\_free\_kicks\_on\_target & Number of direct free-kicks on-target shot by the player \\
        total\_corners  & Number of corners shot by the player  \\
        total\_successful\_penalties & Number of penalties scored by the player  \\
        total\_successful\_linkup\_plays   & Number of linkup plays completed by the player \\
        total\_accelerations & Number of accelerations done by the player \\
        total\_pressing\_duels  & Number of pressing duels attempted by the player  \\
        total\_pressing\_duels\_won  & Number of pressing duels won by the player  \\
        total\_loose\_ball\_duels  & Number of loose ball duels attempted by the player   \\
        total\_loose\_ball\_duels\_won & Number of ball duels won by the player \\
        total\_missed\_balls &  Number of missed balls   \\
        total\_shot\_assists   & Number of shots assists done by the player \\
        total\_shot\_on\_target\_assists & Number of shot on target assists by the player   \\
        total\_recoveries & Number of recoveries by the player \\
        total\_opponent\_half\_recoveries & Number of recoveries in the opponent half by the player \\
        total\_dangerous\_opponent\_half\_recoveries & Number of dangerous recoveries in the opponent half \\
        total\_losses  & Number of ball losses  \\
        total\_own\_half\_losses  & Number of ball losses in own half   \\
        total\_dangerous\_own\_half\_losses & Number of dangerous recoveries in own half  \\
        total\_xg\_shot & Total expected goals from the player's shots \\
        total\_xg\_assist & Total expected goals from the player's passes \\
        total\_xg\_save & Total expected goals saved by the player\\
        total\_received\_pass  & Number of passes received by the player \\
        total\_touch\_in\_box & Number of touches in the box by the player \\
        total\_progressive\_run & Number of progressive run by the player  \\
        total\_offsides & Number of offsides by the player \\
        total\_clearances & Number of clearances by the player   \\
        total\_second\_assists & Number of second assists by the player \\
        total\_third\_assists & Number of third assists by the player \\
        total\_shots\_blocked  & Number of shots blocked by the player  \\
        total\_fouls\_suffered  & Number of fouls in favor of the player\\
        total\_progressive\_passes & Number of progressive passes by the player \\
        total\_counterpressing\_recoveries & Number of counterpressing recoveries by the player \\
        total\_sliding\_tackles & Number of sliding tackles by the player \\
        total\_goal\_kicks & Number of goal kicks by the player \\
        total\_dribbles\_against & Number of dibbles attempted against the player\\
        \bottomrule
        \end{tabular}}
        \vspace{0.5em}
        \caption{Wyscout list of game statistics (part II)}
        \label{tab:wyscoutglos2}
    \end{table}
    
    \begin{table}[ht]
        \centering
        \resizebox{\linewidth}{!}{
        \begin{tabular}{ll}
        \toprule
        {} & \textbf{Explanation} \\
        \midrule
        total\_dribbles\_against\_won & Number of dribbles uncompleted against the player  \\
        total\_goal\_kicks\_short  & Number of short goal kicks attempted by the player\\
        total\_goal\_kicks\_long & Number of long goal kicks attempted by the player  \\
        total\_shots\_on\_target & Number of shots on target by the player  \\
        total\_successful\_progressive\_passes  & Number of successful progressive passes by the player\\
        total\_successful\_sliding\_tackles & Number of sliding tackles completed by the player \\
        total\_successful\_goal\_kicks & Number of goal kicks completed by the player \\
        total\_field\_aerial\_duels & Number of field aerial duels attempted by the player \\
        total\_field\_aerial\_duels\_won & Number of field aerial duels won by the player \\
        total\_gk\_clean\_sheets & Number of clean sheets by the goalkeeper \\
        total\_gk\_conceded\_goals & Number of goals conceded by the goalkeeper  \\
        total\_gk\_shots\_against  &  Number of shots against the goalkeeper \\
        total\_gk\_exits  & Number of goalkeeper exits attempted by the player \\
        total\_gk\_successful\_exits & Number of successful goalkeeper exits completed during the game \\
        total\_gk\_aerial\_duels  & Number of aerial duels attempted by the goalkeeper \\
        total\_gk\_aerial\_duels\_won & Number of aerial duels won by the goalkeeper \\
        total\_gk\_saves   & Number of saves by the goalkeeper  \\
        total\_new\_duels\_won & Number of new duels won \\
        total\_new\_defensive\_duels\_won & Number of new duels new defensive duels won \\
        total\_new\_offensive\_duels\_won & Number of new duels new offensive duels won \\
        total\_new\_successful\_dribbles & Number of new dribbles won \\
        total\_lateral\_passes & Number of lateral passes attempted \\
        total\_successful\_lateral\_passes & Number of lateral passes completed \\
        \bottomrule
        \end{tabular}}
        \vspace{0.5em}
        \caption{Wyscout list of game statistics (part III)}
        \label{tab:wyscoutglos3}
    \end{table}

    \begin{table}[ht]
        \centering
         \resizebox{\linewidth}{!}{
         \begin{tabular}{lll}
        \toprule
        {} &                                      \textbf{Numerator feature} &                                      \textbf{Denominator feature} \\
        \midrule
        ratio\_minutes             &                    total\_minutes\_on\_field &                    total\_matches\\
        ratio\_goals\_shots         &                   total\_goals &                   total\_shots \\
        ratio\_xg\_shots            &                             total\_xg\_shot &                             total\_shots \\
        ratio\_goals\_xg            &                               total\_goals &                               total\_xg\_shot \\
        ratio\_assists\_xa          &                           total\_xg\_assist &                           total\_assists \\
        ratio\_duels\_won           &                           total\_duels\_won &                           total\_duels \\
        ratio\_def\_duels\_won       &                 total\_defensive\_duels\_won &                 total\_defensive\_duels \\
        ratio\_off\_duels\_won       &                 total\_offensive\_duels\_won &                 total\_offensive\_duels \\
        ratio\_air\_duels\_won       &                        total\_aerial\_duels\_won &                        total\_aerial\_duels \\
        ratio\_passes              &                   total\_successful\_passes &                   total\_passes \\
        ratio\_smart\_passes        &             total\_successful\_smart\_passes &             total\_smart\_passes \\
        ratio\_third\_passes        &    total\_successful\_passes\_to\_final\_third &    total\_passes\_to\_final\_third \\
        ratio\_crosses             &                  total\_successful\_crosses &                  total\_crosses \\
        ratio\_for\_passes          &           total\_successful\_forward\_passes &           total\_forward\_passes \\
        ratio\_back\_passes         &              total\_successful\_back\_passes &              total\_back\_passes \\
        ratio\_through\_passes      &           total\_successful\_through\_passes &           total\_through\_passes \\
        ratio\_key\_passes          &               total\_successful\_key\_passes &               total\_key\_passes \\
        ratio\_vert\_passes         &          total\_successful\_vertical\_passes &          total\_vertical\_passes \\
        ratio\_long\_passes         &              total\_successful\_long\_passes &              total\_long\_passes \\
        ratio\_dribbles            &                 total\_successful\_dribbles &                 total\_dribbles \\
        ratio\_def\_actions         &                   total\_defensive\_actions &                   total\_actions \\
        ratio\_att\_actions         &                   total\_attacking\_actions &                   total\_actions \\
        ratio\_penalties           &                           total\_successful\_penalties &                           total\_penalties \\
        ratio\_linup\_plays         &             total\_successful\_linkup\_plays &             total\_linkup\_plays \\
        ratio\_pressing\_duels      &                  total\_pressing\_duels\_won &                  total\_pressing\_duels \\
        ratio\_loose\_ball          &                total\_loose\_ball\_duels\_won &                total\_loose\_ball\_duels \\
        ratio\_opp\_recoveries      &            total\_opponent\_half\_recoveries &            total\_recoveries \\
        ratio\_dang\_recoveries     &  total\_dangerous\_opponent\_half\_recoveries &  total\_opponent\_half\_recoveries \\
        ratio\_own\_losses          &                     total\_own\_half\_losses &                     total\_losses \\
        ratio\_dang\_losses         &           total\_dangerous\_own\_half\_losses &           total\_own\_half\_losses \\
        ratio\_dribbles\_against    &                total\_dribbles\_against\_won &                total\_dribbles\_against \\
        ratio\_field\_aerial\_duels  &                  total\_field\_aerial\_duels\_won &                  total\_field\_aerial\_duels \\
        ratio\_save\_xs            &                             total\_gk\_save &                             total\_xg\_save \\
        ratio\_succ\_exit                &                            total\_gk\_successful\_exits &                            total\_gk\_exits \\
        ratio\_gk\_air\_duels        &                 total\_gk\_aerial\_duels\_won &                 total\_gk\_aerial\_duels \\
        ratio\_lat\_passes          &           total\_successful\_lateral\_passes &           total\_lateral\_passes \\
        ratio\_clean\_game          &                     total\_gk\_clean\_sheets &                     total\_matches \\
        \bottomrule
        \end{tabular}}
        \vspace{0.5em}
        \caption{Engineered ratio features}
        \label{tab:feateng}
    \end{table}

    \begin{table}[!ht]
    \centering
    \begin{minipage}{0.45\textwidth}
    \begin{tabular}{c c}
    \toprule
         \textbf{Rank} &\textbf{Player name}\\ 
         \midrule
        1 & Bukayo Saka \\ 
        2 & Pedri \\ 
        3 & Jude Bellingham \\ 
        4 & Josko Gvardiol \\ 
        5 & Eduardo Camavinga \\ 
        6 & Gavi \\ 
        7 & Jamal Musiala \\ 
        8 & Arnaud Kalimuendo \\ 
        9 & Nuno Mendes \\ 
        10 & Kamaldeen Sulemana \\ 
        11 & Florian Wirtz \\ 
        12 & Nico Williams \\ 
        13 & Ryan Gravenberch \\ 
        14 & Yunus Musah \\ 
        15 & Harvey Elliott \\ 
        16 & Kristjan Asllani \\ 
        17 & Giorgio Scalvini \\ 
        18 & Carney Chukwuemeka \\ 
        19 & Elye Wahi \\ 
        20 & Castello Lukeba \\ 
        21 & Ansu Fati \\ 
        22 & Destiny Udogie \\ 
        23 & Joe Gelhardt \\ 
        24 & Pape Sarr \\ 
        25 & Yan Couto \\ 
        26 & Eliot Matazo \\ 
        27 & Ansgar Knauff \\ 
        28 & Nicola Zalewski \\ 
        29 & Mohamed-Ali Cho \\ 
        30 & Reinier \\ 
        
        \bottomrule
    \end{tabular}
    \end{minipage}
    \begin{minipage}{0.45\textwidth}
    \begin{tabular}{c c}
    \toprule
         \textbf{Rank} &\textbf{Player name}\\ 
         \midrule
        31 & Hugo Ekitike \\ 
        32 & Edoardo Bove \\ 
        33 & Malik Tillman \\ 
        34 & Ki-Jana Hoever \\ 
        35 & Aaron Hickey \\ 
        36 & Noni Madueke \\ 
        37 & Brian Brobbey \\ 
        38 & Jan Thielmann \\ 
        39 & Karim Adeyemi \\ 
        40 & Maarten Vandevoordt \\ 
        41 & Youssoufa Moukoko \\ 
        42 & Lazar Samardzic \\ 
        43 & Mattia Viti \\ 
        44 & Matteo Cancellieri \\ 
        45 & Tomas Suslov \\ 
        46 & Mathys Tel \\ 
        47 & Benjamin Sesko \\ 
        48 & Anouar Ait El Hadj \\ 
        49 & Luka Romero \\ 
        50 & Fabian Rieder \\ 
        51 & Jakub Kaminski \\ 
        52 & Stipe Biuk \\ 
        53 & Luka Sucic \\ 
        54 & Wilfried Gnonto \\ 
        55 & Amar Dedic \\ 
        56 & Sebastiano Esposito \\ 
        57 & Bjorn Meijer \\ 
        58 & Kacper Kozlowski \\ 
        59 & Joelson Fernandes \\ 
        60 & Cher Ndour \\ 
        
        \bottomrule
    \end{tabular}
    \end{minipage}
    \vspace{0.4em}
    \caption{Predicted ranking of Golden Boy nominees}\label{tab:fullranking}
\end{table}